\documentclass[12pt, draftclsnofoot,onecolumn]{IEEEtran}
\usepackage{cite}
\usepackage{graphicx}
\usepackage{amsmath}
\usepackage{amsfonts}
\usepackage{balance}
\usepackage{fancyhdr}
\usepackage{graphicx,times,cite,amssymb,epsfig,amsthm,color}

\newlength{\figwidth}
\normalsize

\usepackage{graphicx}
\ifCLASSINFOpdf
\else
\fi
\usepackage{amsmath}
\DeclareMathOperator*{\argmax}{argmax}
\DeclareMathOperator*{\argmin}{argmin}

\usepackage[cmintegrals]{newtxmath}
\usepackage{algorithm}
\usepackage{algorithmic}

\usepackage{stfloats}
\makeatletter
\newcommand*{\rom}[1]{\expandafter\@slowromancap\romannumeral #1@}
\makeatother

\newcommand{\numfreqs}{L}
\newcommand{\numphases}{M}
\newcommand{\numwaveforms}{M_T}
\newcommand{\indsubpulse}{l}

\begin{document}
%
\title{Combined Radar and Communications \\with Phase-Modulated Frequency Permutations}
%
%
%

\author{Tian~Han,~\IEEEmembership{Student~Member,~IEEE,}
        Rajitha~Senanayake,~\IEEEmembership{Member,~IEEE,}
        Peter~Smith,~\IEEEmembership{Fellow,~IEEE,}
        Jamie~Evans,~\IEEEmembership{Senior~Member,~IEEE,}
        William~Moran,~\IEEEmembership{Member,~IEEE,}
        and~Robin~Evans,~\IEEEmembership{Life~Fellow,~IEEE}


}

\maketitle

\begin{abstract}
This paper focuses on the combined radar and communications problem and conducts a thorough analytical investigation on the effect of phase and frequency change on the communication and sensing functionality. First, we consider the classical stepped frequency radar waveform and modulate data using $M$-ary phase shift keying (MPSK). Two important analytical tools in radar waveform design, namely the ambiguity function (AF) and the Fisher information matrix (FIM) are derived, based on which, we make the important conclusion that MPSK modulation has a negligible effect on radar local accuracy. Next, we extend the analysis to incorporate frequency permutations and propose a new signalling scheme in which the mapping between incoming data and waveforms is performed based on an efficient combinatorial transform called the Lehmer code. We also provide an efficient communications receiver based on the Hungarian algorithm. From the communications perspective, we consider the optimal maximum likelihood (ML) detector and derive the union bound and nearest neighbour approximation on the block error probability. From the radar sensing perspective, we discuss the broader structure of the waveform based on the AF derivation and quantify the radar local accuracy based on the FIM. Extensive numerical examples are provided to illustrate the accuracy of our results. 

\end{abstract}


%
\IEEEpeerreviewmaketitle

\section{Introduction}
%
%
%
%

\IEEEPARstart{T}{he} integration of radar sensing and communication is a promising design paradigm in which sensing and communication functionalities share the same hardware and spectrum resources. Traditionally, communications and radar sensing were designed separately to focus on domain specific challenges. However, as communication systems start to use the millimeter-wave (mmWave) frequency band which is traditionally used in radar, there has been an increasing amount of research interest in the integration of the two functionalities \cite{Hassanien16, Sturm11, Zheng17, Liu20, zhang2021, liu2021, Tan21, Liu22vehicular1}. Such joint systems can contribute to reducing the system cost and power consumption, as well as alleviating concerns for spectrum congestion \cite{Ma20}. 

\subsection{Related Works} \label{sec:related_works}
A key aspect of the co-existence of these two functions in an integrated system is the design of a joint waveform that is capable of transmitting information and performing radar sensing simultaneously. This convergence could be achieved based on different approaches as discussed in \cite{Liu20, zhang2021, liu2021, Tan21, Liu22vehicular2, Oliveira22, Thoma21}. 
For example, some research focuses on embedding information into traditional radar waveforms for joint application. 
In \cite{Blunt10}, the traditional linear frequency modulated (LFM) waveform is considered for data transmission. However, the communication symbol rate corresponds to the chirp rate only, which is much lower than the symbol rate that can be achieved by a communication system with the same bandwidth. 
In \cite{McCormick2019}, a phase-attached radar-communications (PARC) framework is extended to a frequency modulated continuous-wave (FMCW) implementation as a means to realise both functions. Stretch processing is employed such that large bandwidth radar operation and fine range resolution can be achieved. In \cite{Sahin17}, information sequences are implemented using continuous phase modulation (CPM) and phase-attached to a polyphase-coded frequency-modulated (PCFM) fixed radar waveform. The adjustable parameters provide control of the range side-lobe modulation (RSM) by trading off bit error rate (BER) and/or data throughput.
Frequency-hopping (FH) multiple-input multiple-output (MIMO) radar is recently introduced to dual-function radar-communication (DFRC) which increase the symbol rate to multiples of pulse repetition frequency \cite{Wu22, Baxter22, Wu21, Eedara18}. In \cite{Baxter22}, a DFRC framework which unifies various existing signalling strategies in FH-MIMO radar systems is presented. Hybrid information embedding schemes are considered in order to further boost the communication data rates. In \cite{Wu21}, a unique FH-MIMO waveform based DFRC scheme is proposed in order to address issues including channel estimation and synchronisation in multi-path channels.

Research is also underway on the opposite design pattern, i.e., using traditional communication waveforms for radar sensing. The use of the orthogonal frequency division multiplexing (OFDM) waveforms for the joint system was considered in \cite{levanon2000multifrequency,Lellouch15,Shi18}. In \cite{Lellouch15}, a wideband OFDM based waveform is presented and the high range resolution processing is derived. Furthermore, the impact of Doppler modulation on the processing is inspected to give recommendations of the OFDM parameters. 
In \cite{Shi18}, the design of a power minimisation-based robust OFDM radar waveform is considered. However, due to the large number of subcarriers in the OFDM waveform it introduces high peak-to-average power ratio (PAPR). This makes the approach challenging for typical radar operation since it demands for power amplifiers with a large linear range. 
Taking a different approach, in \cite{Kumari15}, the IEEE 802.11ad-based waveform which is traditionally used for wireless local area networks (WLANs) is proposed for long-range radar (LRR) applications at the 60 GHz unlicensed band. The preamble of this waveform, which consists of Golay complementary sequences with good correlation properties, is exploited to make it suitable for radar applications. In \cite{Kumari20}, a virtual waveform design based on IEEE 802.11ad is proposed for an adaptive mmWave joint communications and radar (JCR) system. The system transmits a few non-uniformly placed preambles to enhance the velocity estimation accuracy, at the cost of a small reduction in the communication data rate. In \cite{Tang21}, a self-interference-resistant IEEE 802.11ad based JCR framework is developed. The Golay sequences and a novel pilot signal design are leveraged to combat self-interference in different sensing scenarios.  

\subsection{Motivations and Contributions}
Despite the efforts made in these designs for JCR systems, many fundamental research problems remain open. In this paper, we are motivated to conduct a rigorous theoretical analysis on the effect of phase and frequency modulation on the radar sensing functionality, which to the best of our knowledge is an important open research problem.  

One of the main challenges of embedding information into conventional radar waveforms is the limited rate of communication due to the lack of randomness. In addition, the radar estimation accuracy is sensitive to some parameters which might be varied during modulation and could cause degradation of radar performance. Taking a different approach to the aforementioned papers, here, we focus on embedding data into the conventional stepped frequency radar waveform by randomising the parameters that have little effect on radar performance. 


Stepped frequency waveforms are commonly used in radar applications. A stepped frequency waveform has multiple subpulses, each with a constant frequency tone, but the frequencies of different subpulses are equally separated by a constant step $\Delta f$ Hz. Examples of stepped frequency radar waveforms include the linear stepped frequency waveform \cite{richards14} and Costas coded waveform \cite{Levanon2004}. 
Phase coding is one of the early methods for pulse compression in radar applications \cite{Levanon2004}. Chirp like polyphase codes, such as Frank codes \cite{frank1963}, have properties of low autocorrelation side-lobe levels and good Doppler tolerance. In \cite{Lewis1981}, modified versions of Frank codes, namely P1 and P2 codes, are proposed. Compared to Frank codes, P1 and P2 codes are more tolerant of receiver band-limiting prior to pulse compression. All of these waveforms, however, maintain a strict sequence of  phase values carefully designed to achieve the best radar performance, which are not suitable for joint operation due to the lack of randomness. 

In this paper, we change the phase and frequency of the conventional stepped frequency radar waveform to form a new waveform that is suitable for both data transmission and radar sensing. First we focus on the phase change by incorporating phase shift keying (PSK) modulation to the classical stepped frequency radar waveform. Recently, in \cite{Eedara18}, an orthogonal FH-MIMO radar waveform with PSK signalling is proposed, which shows that the highest side-lobe level of the ambiguity function (AF) is decreased due to phase modulation. However, the conclusion in this paper was drawn based on numerical simulations. Next, we extend this by adopting the frequency permutation based modulation introduced in \cite{Senanayake21VTC, Senanayake22TWC} and exploiting the randomness of the permutation selection on top of the phase modulation to formulate a new modulation scheme. 
The joint use of both phase and frequency, solely for the communication purposes, has recently been analysed in \cite{kundu20}. They proposed a novel $M$-ary frequency-phase keying modulation, in which the frequency and phase are operated in a discrete, slot-by-slot manner. Similar to \cite{kundu20}, in our proposed modulation method frequency is changed in a discrete manner, based on the selected permutation while the phase is changed based on PSK modulation. This allows us to use phase and frequency independently to modulate data and as a result transmit more data between nodes. 
We provide a rigorous theoretical analysis on the proposed waveform, both in-terms of communication performance and radar performance, and make comparisons with other proposed approaches such as \cite{Senanayake22TWC}, which only consider frequency modulation.  
The main contributions of this paper are as follows:
\begin{itemize}
\item First, we analyse the effect of phase modulation on the radar performance of the linear stepped frequency waveform. 
More specifically, we derive the AF and discuss its broader structure and side-lobe levels. We also derive approximate expressions for the Cramer-Rao lower bounds (CRLBs) on the delay and Doppler estimation errors based on the Fisher information matrix (FIM), and provide a detailed evaluation of the local accuracy. Based on the analytic tools, we analyse the loss in radar estimation accuracy when incorporating communications functionality into the traditional linear stepped frequency waveform.

\item Next, we combine phase modulation and permutation of the frequency tones to propose a novel integrated radar and communication waveform. We present a new signalling scheme which allows more data to be transmitted and provide an efficient implementation for the mapping between the incoming data and the corresponding waveform based on a combinatorial transform called the Lehmer code. 
We derive the AF, the FIM and approximations to the CRLBs, based on which the cost of embedding data using phase modulation is analysed. We make a performance comparison to the work in \cite{Senanayake22TWC} which, different to our paper, only uses frequency permutations for the data modulation.

\item From a communications perspective, we consider maximum likelihood (ML) detection and analyse the error probability of the receiver in different wireless communication models. As we proposed a novel modulation scheme, the derivation of its error probability performance is new and challenging. For additive white Gaussian noise (AWGN) channels and correlated Rician fading channels, we derive the union bound as well as the nearest neighbour approximation on the block error probability. In addition, we derive a new upper bound on the block error probability under the correlated Rayleigh fading model. To deal with the high complexity of an exhaustive search in ML detection, we provide an efficient implementation of the optimal ML receiver based on a modified Hungarian algorithm. 
\end{itemize}
Extensive numerical examples are provided to illustrate the accuracy of our results.

The remainder of this paper is organised as follows. In Section \ref{sec:problem_formulation}, we formulate the problem and describe the joint radar and communication system. The use of specific performance measures for both radar and communication is also motivated. The phase modulated linear stepped frequency waveform and a novel frequency permutation and phase modulation based waveform are introduced in Section \ref{sec:pm_step} and \ref{sec:pm_perm}, respectively. The radar performance of the two waveforms is analysed and numerical examples are provided to support the analysis. The communication performance of the frequency permutation and phase modulation based waveform is analysed in Section \ref{sec:comms_performance} along with numerical examples. In Section \ref{sec:conclusion}, the conclusions of this paper as well as some possible future extensions are provided.

\section{Problem Formulation} \label{sec:problem_formulation}
\subsection{System Model}
\begin{figure}[t]
\centering
\includegraphics[width=2.5in]{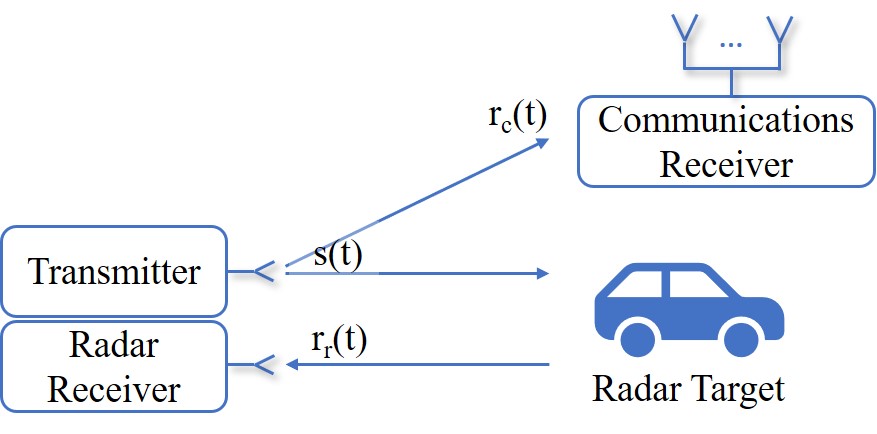}
\caption{The JCR system model.}
\label{fig:system}
\end{figure}

We consider a JCR system model, as illustrated in Fig. \ref{fig:system}, where the transmitter sends a common waveform to both the radar target and the communications receiver. The signal reflected from the radar target is received and processed by the radar receiver, which is co-located with the transmitter, to estimate the range and relative velocity of the target. The signal received at the communications receiver is processed to detect the transmitted information.

First, let us focus on the classical stepped frequency radar waveform with linearly increasing frequency tones. The complex envelope of such a stepped frequency radar waveform with $\numfreqs$ subpulses of length $T$ can be given by 
\begin{equation}
    s(t) = \sqrt{\frac{E}{\numfreqs T}}\sum_{\indsubpulse=0}^{\numfreqs-1}s_p(t-\indsubpulse T)\exp{(j2\pi f_\indsubpulse(t-\indsubpulse T))}, \label{eq:signal_stepfreq}
\end{equation}
where the energy of the waveform
$E = \int_0^{\numfreqs T} |s(t)|^2 dt$
and $s_p(t)$ is a simple rectangular pulse
\begin{equation*}
\begin{aligned}
s_p(t) = \left\{\begin{aligned}
& 1, \quad 0\leq t \leq T \\
& 0, \quad \text{otherwise}.
\end{aligned}
\right.
\end{aligned}
\end{equation*}
We also assume that the frequency tones are in ascending order and the difference between two successive frequency tones is $\Delta f = n/T$, where $n$ is a positive integer. As such,  the frequencies are orthogonal to each other and the frequency tone of the $\indsubpulse$-th subpulse can be expressed as 
\begin{equation}
    f_\indsubpulse = f_0 + n(\indsubpulse-1)/T. \label{eq:freqtones}
\end{equation}
In \eqref{eq:signal_stepfreq}, there are several parameters including the amplitude, phase and frequency of the waveform that can be changed in order to modulate data. In the current paper, we investigate novel methods of modulating the phase and frequency without causing significant changes to radar sensing functionality. We keep the amplitude fixed because constant amplitude is extremely important for the efficient use of the power amplifiers in radar operation \cite{richards14}. 

\subsection{Radar Performance Measures}
In this paper we focus on AF and FIM, the two key analytical  tools used in radar waveform design. For convenience we have provided the definitions of these tools 
as follows.  

The AF of a waveform describes the output of a matched filter when the signal input to the radar receiver is delayed and Doppler shifted by certain amounts. 
Based on \cite[eq.(4.30)]{richards14}, the complex ambiguity function of the waveform $s(t)$ in \eqref{eq:signal_stepfreq} can be expressed as
\begin{equation}
    \hat{A}(\tau, \omega) = \int_{-\infty}^{\infty} s(t) s^*(t-\tau)e^{j\omega t} d t, \label{eq:cAF}
\end{equation}
where $\tau$ is the time delay to the expected matched filter peak output, and $\omega$ is the Doppler mismatch between the Doppler shift of the received signal and that for which the filter is designed. Then the AF is defined as the magnitude of the complex ambiguity function and is given by 
\begin{equation}
\begin{aligned}
A(\tau, \omega) = \left| \hat{A}(\tau, \omega) \right|. \label{eq:AF}
\end{aligned}
\end{equation}
The AF provides a measure of the degree of similarity between the transmitted waveform and its time and frequency shifted version \cite{vantrees01}. The local accuracy can be analysed based on the shape of the AF around the origin, while the ambiguity can be studied by the AF side-lobe levels.

The FIM for delay and Doppler shift estimations provide a more detailed evaluation of the local accuracy of the waveform \cite{vantrees01}.
Consider the complex envelope of the radar received signal reflected by a single target, which can be written as
\begin{equation}
r_r(t) = \Tilde{b} s(t-\tau) e^{j\omega t} + n(t),
\end{equation}
where $\tau$ and $\omega$ are the target delay and Doppler shift to be estimated, $\Tilde{b}$ is a complex Gaussian random variable with zero mean and unit variance and $n(t)$ represents the baseband complex additive white Gaussian noise process with zero mean and power spectrum density $N_0$ \cite[eq.(10.5)]{vantrees01}. The Rayleigh fading variable, $\Tilde{b}$, models the reflection from a target with multiple reflecting surfaces.
The FIM for $\tau$ and $\omega$ is defined as
\begin{equation} \label{eq:FIM}
\begin{aligned}
\boldsymbol{J} = \left[
 \begin{array}{cc}
  J_{11} & J_{12} \\
  J_{21} & J_{22} \end{array} \right],
\end{aligned}
\end{equation}
where subscript 1 denotes $\tau$ and subscript 2 denotes $\omega$. Using \cite[eq.(10.63)-(10.65)]{vantrees01}, the elements of $\boldsymbol{J}$ can be written as
\begin{equation}
    J_{11} = C[\overline{\omega^2}-\Bar{\omega}^2],
\end{equation}
\begin{equation}
    J_{12} = J_{21}  =  C[\overline{\omega \tau}-\Bar{\omega}\Bar{\tau}],
\end{equation}
\begin{equation}
    J_{22} = C[\overline{\tau^2}-\Bar{\tau}^2],
\end{equation}
where 
    $ C = \frac{2}{N_0(1+N_0)}, $
$~\overline{\omega^2}$, $\overline{\omega \tau}$ and $\overline{\tau^2}$ are given in \cite[eq.(10.67)-(10.69)]{vantrees01}, respectively. Note that $\Bar{\omega}$ and $\Bar{\tau}$ can be calculated by replacing $\omega^2$ with $\omega$ in \cite[eq.(10.67)]{vantrees01} and replacing $u^2$ by $u$ in \cite[eq.(10.69)]{vantrees01}, respectively.
Based on the FIM elements we can derive the CRLBs on the delay and Doppler estimation errors.  




\section{Phase Modulated Stepped Frequency Waveform} \label{sec:pm_step}
In this section, we introduce phase modulation to embed data into the waveform in \eqref{eq:signal_stepfreq}. Suppose \numphases PSK is used independently in each subpulse. 
 Then, the waveform can be expressed as
\begin{equation}
s(t) = \sqrt{\frac{E}{\numfreqs T}}\sum_{\indsubpulse=0}^{\numfreqs-1}s_p(t-\indsubpulse T)\exp{(j(2\pi f_\indsubpulse(t-\indsubpulse T)+\theta_\indsubpulse))}, \label{eq:signal_steppsk}
\end{equation}
where $\theta_\indsubpulse \in \{0, 2\pi/\numphases, ... , 2\pi(\numphases-1)/\numphases \}$ denotes the phase modulated into the $\indsubpulse$-th subpulse of the waveform. 
While adding \numphases PSK allows us to send data, it impacts the radar sensing functionality. Thus, we proceed to analyse the AF and the FIM to discuss the effect of phase change on the radar performance.   

\subsection{Ambiguity Function} \label{sec:AF_step}
In order to derive the AF, we normalise the signal energy to one and re-express the waveform in (\ref{eq:signal_steppsk}) as
\begin{equation}
    s(t) = \sqrt{\frac{1}{\numfreqs T}}\sum_{\indsubpulse=0}^{M-1}s_p(t-\indsubpulse T)\exp{(j(\omega_\indsubpulse (t-\indsubpulse T) + \theta_\indsubpulse))},  \label{eq:signal_steppsk_radar}
\end{equation}
where $\omega_\indsubpulse  = 2 \pi f_\indsubpulse$ is the frequency in radians per second. 
Substituting (\ref{eq:signal_steppsk_radar}) into (\ref{eq:cAF}) we can write the complex AF of $s(t)$ as 
\begin{equation}
\hat{A}(\tau, \omega) = \frac{1}{\numfreqs T}\sum_{\indsubpulse =0}^{\numfreqs-1}\sum_{n=0}^{\numfreqs-1}\int_{-\infty}^{\infty}s_p(s-\indsubpulse T)s^*_p(s-nT-\tau) e^{(j(\omega s + \omega_\indsubpulse (t-\indsubpulse T)-\omega_n(s-\tau-nT)+\theta_\indsubpulse -\theta_n))} ds.
\label{eq:cAF_steppsk1}
\end{equation}
Writing the complex AF of 
$s_p(t)$ as $\hat{A}_p(\tau, \omega)$ and rearranging (\ref{eq:cAF_steppsk1}) we obtain
\begin{equation}
\hat{A}(\tau, \omega) = \frac{1}{\numfreqs T}\sum_{\indsubpulse =0}^{\numfreqs-1}\sum_{n=0}^{\numfreqs-1}\hat{A}_p(\tau+(n-\indsubpulse )T, \omega - (\omega_n-\omega_\indsubpulse ))  e^{j(\omega \indsubpulse T+\omega_n(\tau+(n-\indsubpulse )T)+(\theta_\indsubpulse -\theta_n))}, \label{eq:cAF_steppsk}
\end{equation}
where $\hat{A}_p(\tau, \omega)$ can be written as
\begin{equation}
\begin{aligned}
\hat{A}_p(\tau,\omega) = \left\{\begin{aligned}
&\frac{e^{j\omega(\tau+T)}-1}{j\omega} , \quad &-T < \tau < 0 \\
& \frac{e^{j\omega T}-e^{j\omega \tau}}{j\omega}, \quad  &0 \leq \tau < T\\
&0, \quad &\text{otherwise}.
\end{aligned}
\right.   \label{eq:spcAF}
\end{aligned}
\end{equation}
Next, by taking the magnitude of (\ref{eq:cAF_steppsk}), we can write the AF of the waveform in (\ref{eq:signal_steppsk_radar}) as
\begin{equation}
A(\tau, \omega) = \left| \frac{1}{\numfreqs T}\sum_{\indsubpulse =0}^{\numfreqs-1}\sum_{n=0}^{\numfreqs-1}\hat{A}_p(\tau+(n-\indsubpulse )T, \omega - (\omega_n-\omega_\indsubpulse ))  e^{j(\omega \indsubpulse T+\omega_n(\tau+(n-\indsubpulse )T)+(\theta_\indsubpulse -\theta_n))} \right|. \label{eq:AF_steppsk}
\end{equation}
The zero-Doppler cut of the AF describes the matched filter output when these is only time delay and no Doppler mismatch. By letting $\omega=0$ in (\ref{eq:AF_steppsk}) we can obtain the zero-Doppler cut as
\begin{equation}
A(\tau, 0) = \left| \frac{1}{\numfreqs T} \sum_{\indsubpulse =0}^{\numfreqs-1}\sum_{n=0}^{\numfreqs-1}\hat{A}_p(\tau+(n-\indsubpulse )T, \omega_\indsubpulse -\omega_n)  e^{j(\omega_n(\tau-(n-\indsubpulse )T)+(\theta_\indsubpulse -\theta_n))} \right|. \label{eq:AF0dop_steppsk}
\end{equation}
Similarly, the zero-delay cut of the AF describes the matched filter output when there is no time delay, which can be found by letting $\tau=0$ in (\ref{eq:AF_steppsk}) to derive
\begin{equation}
A(0, \omega) = \left| \frac{1}{\numfreqs T} \sum_{\indsubpulse =0}^{\numfreqs-1}\sum_{n=0}^{\numfreqs-1}\hat{A}_p((n-\indsubpulse )T, \omega - (\omega_n-\omega_\indsubpulse )) e^{j(\omega \indsubpulse T+\omega_n(n-\indsubpulse )T+(\theta_\indsubpulse -\theta_n))} \right|.  \label{eq:AF0del_steppsk}
\end{equation}
Due to the fact that $\hat{A}_p(\tau,\omega)$ in (\ref{eq:spcAF}) is non-zero only when $-T<\tau<T$, it is obvious that $\hat{A}_p((n-\indsubpulse )T, \omega - (\omega_n-\omega_\indsubpulse ))$ is non-zero only when $n=\indsubpulse $. Therefore, the zero-delay cut in \eqref{eq:AF0del_steppsk} can be simplified as
\begin{equation}
A(0,\omega) = \left| \frac{1}{\numfreqs T}\sum_{\indsubpulse =0}^{\numfreqs-1}\hat{A}_p(0, \omega)e^{j\omega \indsubpulse T} \right|, \label{eq:AF0delsimp_steppsk}
\end{equation}
which indicates that the zero-delay cut does not change with the phase modulation. 

\subsection{Fisher Information Matrix and Cramer-Rao Lower Bounds} \label{sec:crlb_step}
Based on the definition of the FIM in \eqref{eq:FIM} we proceed to derive the elements $J_{11}$, $J_{12}$, $J_{21}$ and $J_{22}$ as follows
\begin{equation}
    J_{11} \approx \frac{2CB}{\numfreqs T}\left( \numfreqs -\sum_{\indsubpulse =0}^{\numfreqs-2}\cos(\omega_0  T + \theta_\indsubpulse -\theta_{\indsubpulse +1}) \right), \label{eq:J11_step}
\end{equation}
\begin{equation}
    J_{12} = J_{21} \approx -\frac{CT^2}{2}\sum_{\indsubpulse =0}^{\numfreqs-1}\omega_\indsubpulse (2\indsubpulse +1), \label{eq:J12_step}
\end{equation}
\begin{equation}
    J_{22} = \frac{C\numfreqs^2T^2}{12}, \label{eq:J22_step}
\end{equation}
where $B$ is the finite bandwidth occupied by the rectangular pulse shaping function $s_p(t)$. 
Detailed derivation of $J_{11}$, $J_{22}$ and $J_{12}$ is not included in the present paper due to page limitations, but the approach follows similar steps as in \cite[Appendix~C, D and E]{senanayake21arxiv}. 
Note that the finite bandwidth $B$ is introduced since $\overline{\omega^2}$ and $\Bar{\omega}^2$ do not converge when $s_p(t)$ is a perfect rectangular pulse. The approximation in \eqref{eq:J11_step} is derived by neglecting small high order terms in the integrals which vanish when $BT \to \infty$ \cite{senanayake21arxiv}.

The FIM is useful to bound the variance of the individual errors. More specifically, when unbiased estimators are used, the variances of delay and Doppler shift estimation errors are lower bounded by the diagonal elements in $\boldsymbol{J}^{-1}$ \cite{vantrees01}. Denoting the CRLBs of the delay estimation and Doppler shift estimation errors as $\text{CRLB}_{\tau}$ and $\text{CRLB}_{\omega}$, respectively, their approximate expressions can be expressed as
 \begin{equation}
\text{CRLB}_{\tau} \approx \frac{C^{-1}\numfreqs^2T}{2\numfreqs B\left(\numfreqs-\sum_{\indsubpulse =0}^{\numfreqs-2}\cos{(\omega_0 T+\theta_\indsubpulse -\theta_{\indsubpulse +1})}\right)-3T^3\left(\sum_{\indsubpulse =0}^{\numfreqs-1}(2\indsubpulse +1)\omega_\indsubpulse  \right)^2}, \label{eq:crlb_t_step}
\end{equation}

\begin{equation}
\text{CRLB}_{\omega} \approx \frac{24C^{-1}B\left(\numfreqs-\sum_{\indsubpulse =0}^{\numfreqs-2}\cos{(\omega_0 T+\theta_\indsubpulse -\theta_{\indsubpulse +1})}\right)}{2\numfreqs^2T^2B\left(\numfreqs-\sum_{\indsubpulse =0}^{\numfreqs-2}\cos{(\omega_0 T+\theta_\indsubpulse -\theta_{\indsubpulse +1})}\right)-3\numfreqs T^5\left(\sum_{\indsubpulse =0}^{\numfreqs-1}(2\indsubpulse +1)\omega_\indsubpulse  \right)^2}. \label{eq:crlb_w_step}
 \end{equation}

\noindent
Note that (\ref{eq:crlb_t_step}) and (\ref{eq:crlb_w_step}) can be further simplified based on \cite[eq.(10.94), (10.95)]{vantrees01} to produce
\begin{equation}
\text{CRLB}_{\tau} \approx \frac{C^{-1}\numfreqs T}{2B\left( \numfreqs -\sum_{\indsubpulse =0}^{\numfreqs-2}\cos(\omega_0 T + \theta_\indsubpulse -\theta_{\indsubpulse +1}) \right)}, \label{eq:crlb_st_step}
\end{equation}
\begin{equation}
\text{CRLB}_{\omega} \approx \frac{12C^{-1}}{\numfreqs^2T^2}. \label{eq:crlb_sw_step}
\end{equation}
While the simplified expressions in (\ref{eq:crlb_st_step}) and (\ref{eq:crlb_sw_step}) provide a looser bound compared to (\ref{eq:crlb_t_step}) and (\ref{eq:crlb_w_step}), they clearly show the effect of parameters on the estimation errors. The delay estimation error lower bound in (\ref{eq:crlb_st_step}) is inversely proportional to the product of the finite bandwidth $B$ and the effective bandwidth $1/T$ of a single subpulse. The Doppler shift estimation error lower bound in (\ref{eq:crlb_sw_step}) is inversely proportional to the square of the time duration of the whole waveform $\numfreqs T$.
To observe the effect of embedding data on local accuracy, we first derive the CRLBs for stepped frequency waveforms without phase modulation by setting $\theta_m=0$ in (\ref{eq:crlb_st_step}) and (\ref{eq:crlb_sw_step}), which results in
\begin{equation}
\text{CRLB}_{\tau} \approx \frac{C^{-1}\numfreqs T}{2B\left( \numfreqs -(\numfreqs-1)\cos(\omega_0 T) \right)}, \label{eq:crlb_stept}
\end{equation}
\begin{equation}
\text{CRLB}_{\omega} \approx \frac{12C^{-1}}{\numfreqs^2T^2}. \label{eq:crlb_stepw}
\end{equation}

Comparing the CRLBs in (\ref{eq:crlb_stept}) and (\ref{eq:crlb_stepw}) with those in (\ref{eq:crlb_st_step}) and (\ref{eq:crlb_sw_step}), it can be found that the phase modulation has exactly no effect on the simplified $\text{CRLB}_{\omega}$ 
while it has some influence on $\text{CRLB}_{\tau}$. Nevertheless, the maximum value (when all $\cos{(\cdot)}$ terms equal $1$) and minimum value (when all $\cos{(\cdot)}$ terms equal $-1$) of the simplified $\text{CRLB}_{\tau}$ are kept unchanged when phase modulation is introduced. 
The discussions above are further analysed using numerical examples in Section \ref{sec:num_step_radar} below.

\subsection{Numerical Examples} \label{sec:num_step_radar}
\begin{figure*}
    \centerline{\includegraphics[width=17.3cm,height=6.7cm]{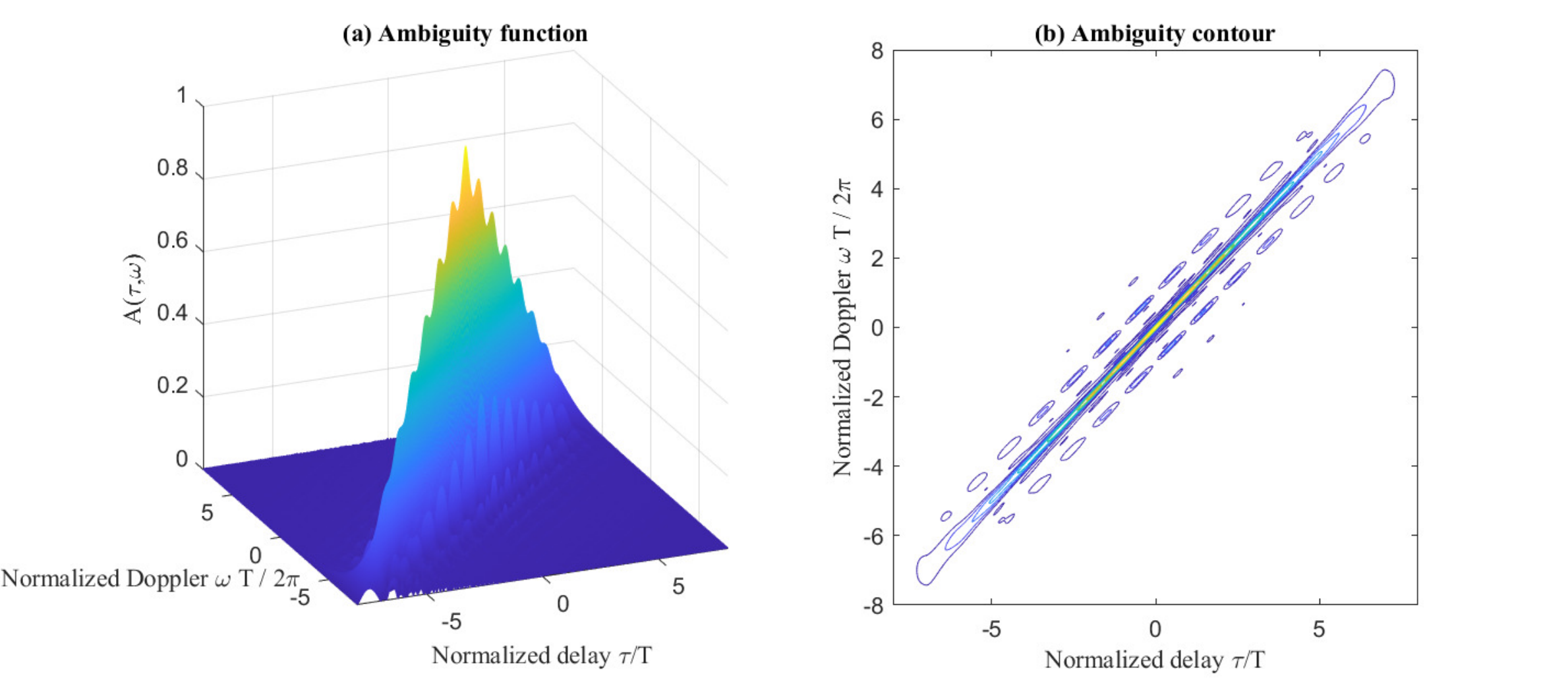}}
    \caption{The (a) three-dimensional surface plot and (b) contour plot of the AF  of a linear stepped frequency waveform for $\numfreqs = 8$. The first frequency $f_0 = 0$ Hz and the step $\Delta f = 1/T$Hz.}\label{fig:AF_step}
\end{figure*}

\begin{figure*}
    \centerline{\includegraphics[width=17.3cm,height=6.7cm]{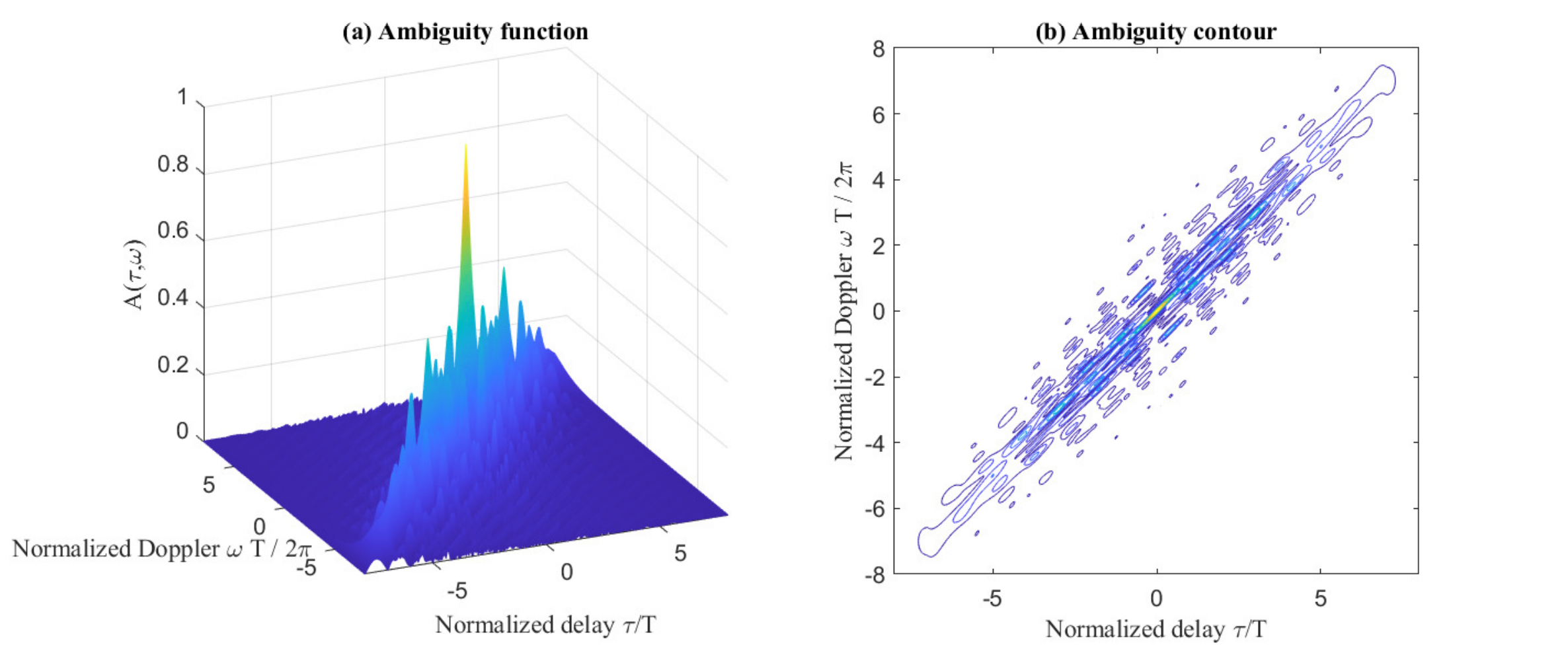}}
    \caption{The (a) three-dimensional surface plot and (b) contour plot of the AF of a linear stepped frequency and PSK based waveform for $\numfreqs = 8$ and $\numphases = 4$. The first frequency $f_0 = 0$ Hz and the step $\Delta f = 1/T$Hz. The phase sequence is $[3\pi/4,3\pi/4,3\pi/4,\pi/2,3\pi/4,\pi/2,\pi/4,\pi/2]$.
    }\label{fig:AF_step_psk}
\end{figure*}

\begin{figure*}
    \centerline{\includegraphics[width=19.5cm,height=7cm]{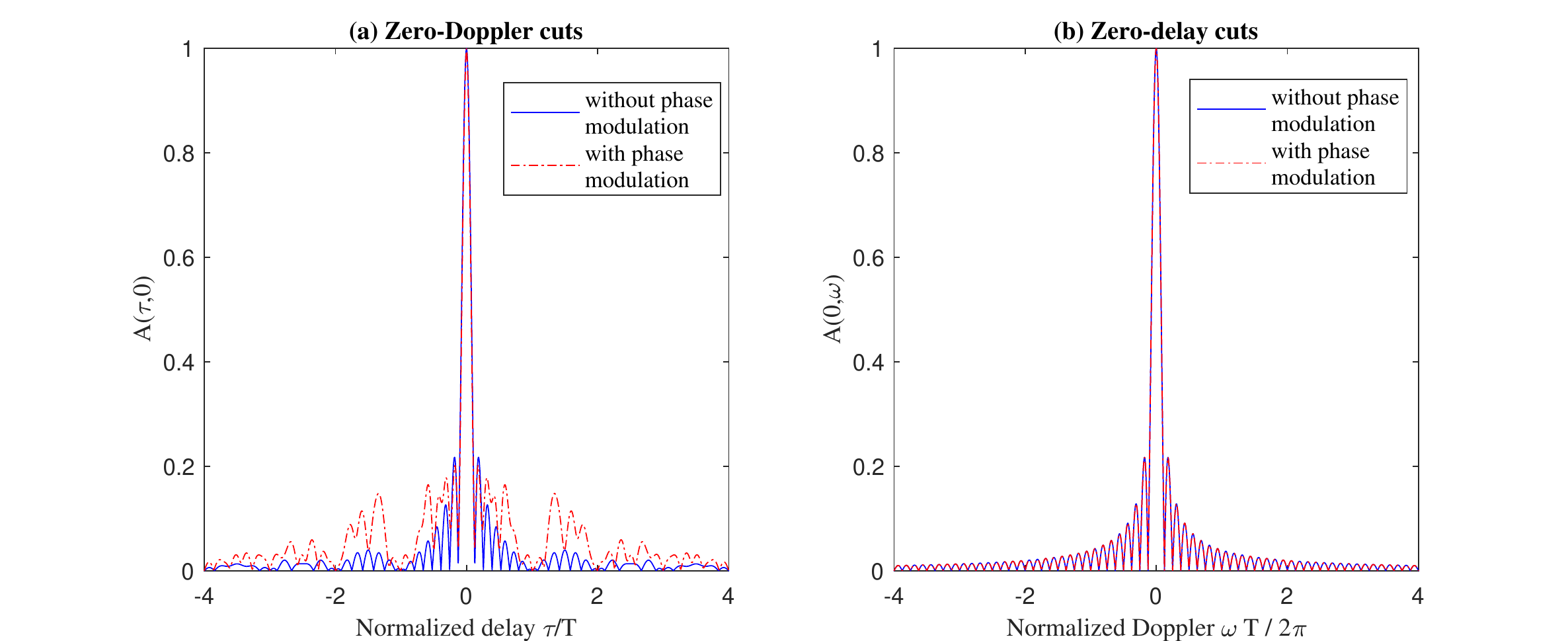}}
    \caption{The (a) zero-Doppler cuts and (b) zero-delay cuts of the AFs in Fig. \ref{fig:AF_step} and Fig. \ref{fig:AF_step_psk}. 
    }\label{fig:0cuts_step}
\end{figure*}

\begin{figure*} 
    \centerline{\includegraphics[width=19.5cm,height=7.5cm]{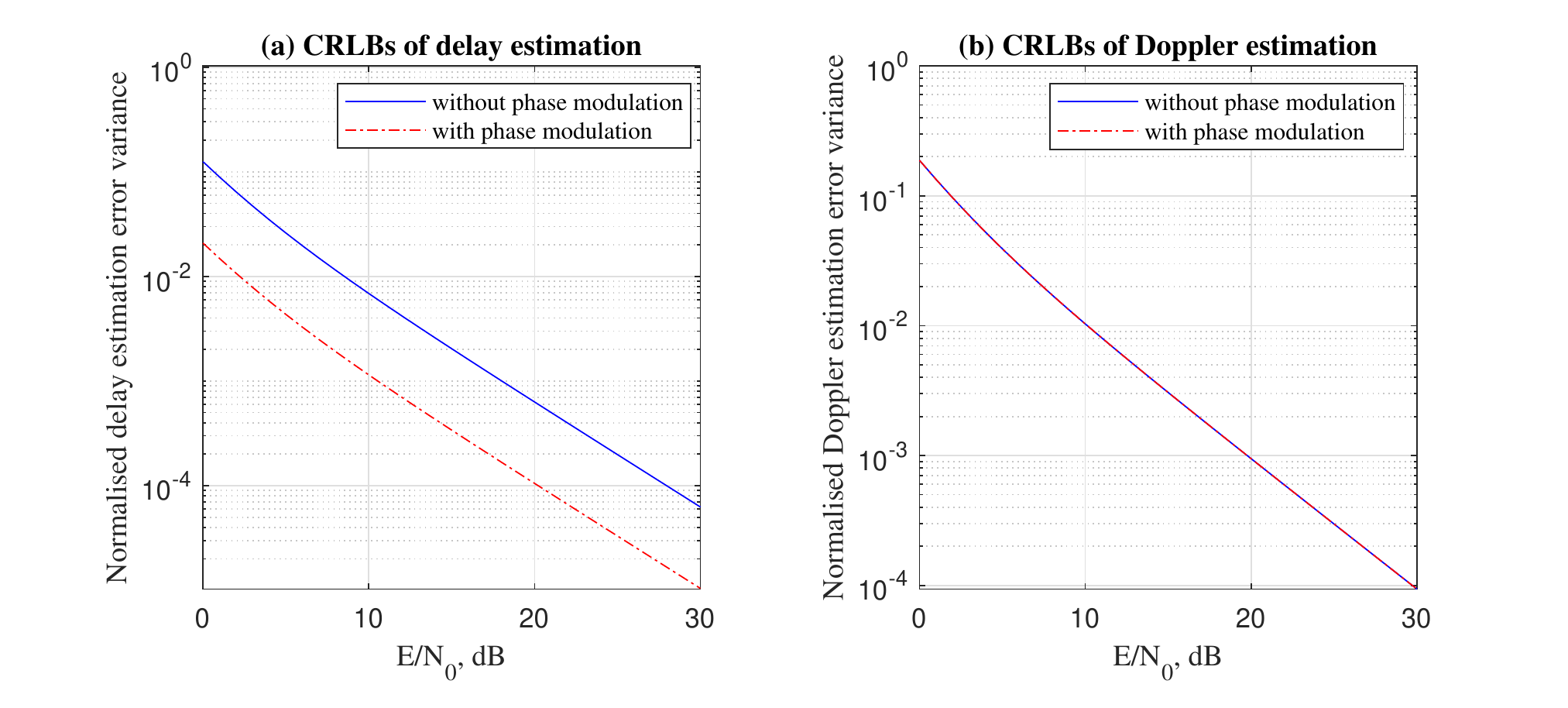}}
    \caption{{The (a) approximated CRLBs in \eqref{eq:crlb_st_step} and \eqref{eq:crlb_stept} on delay estimation errors and (b) approximated CRLBs in \eqref{eq:crlb_sw_step} and \eqref{eq:crlb_stepw} on Doppler shift estimation errors for stepped frequency based waveforms with and without phase modulation.}. 
    }\label{fig:CRLB_step}
\end{figure*}

In this section, we provide numerical examples to support the radar performance analyses in Section \ref{sec:AF_step} and \ref{sec:crlb_step}. Fig. \ref{fig:AF_step} to Fig. \ref{fig:0cuts_step} illustrate the AF while Fig. \ref{fig:CRLB_step} illustrates the CRLBs on delay and Doppler shift estimation errors.

Fig. \ref{fig:AF_step} and Fig. \ref{fig:AF_step_psk} plot the AFs and the corresponding contour plots of the stepped frequency based waveforms without phase modulation and with phase modulation, respectively. The number of frequency tones  $\numfreqs = 8$. In Fig. \ref{fig:AF_step_psk}, we consider quadrature phase shift keying (QPSK) modulation and set $M = 4$. We have given the plot for one random sequence of phases and this particular phase sequence will be used in all the plots for the phase modulated waveform in the remainder of this section. From Fig. \ref{fig:AF_step} and Fig. \ref{fig:AF_step_psk} we observe that the broader structure of the AF is slightly changed due to the phase modulation. To be more specific, when there is no phase modulation, most of the volume under the AF is concentrated in a ridge located in the first and third quadrants. Although there still exists a ridge when phase modulation is introduced, the volume is slightly spread out. However, the AF around the origin is almost unaffected by the phase modulation.

To gain more insight into the AF and see the behaviour around the origin more clearly, Fig. \ref{fig:0cuts_step} provides one dimensional cuts of the AFs in Fig. \ref{fig:AF_step} and Fig. \ref{fig:AF_step_psk}. From Fig. \ref{fig:0cuts_step} (a) we observe that the zero-Doppler cut is changed due to the phase modulation but the change of the highest side-lobe level as well as the curvature at the origin is negligible, which indicates that phase modulation has little impact on the delay estimation accuracy. 
From Fig. \ref{fig:0cuts_step} (b) we observe that the zero-delay cuts of the two AFs are exactly the same, which indicates that the phases of subpulses have no impact on the zero-delay cut. Hence, there is very little impact on the Doppler shift estimation accuracy. This also agrees with the theoretical results in \eqref{eq:AF0delsimp_steppsk} as well as the analysis in \eqref{eq:crlb_sw_step}.

Next, we focus on quantitative performance measurements of the delay and Doppler shift estimation accuracies, which can be illustrated by the CRLBs. Fig. \ref{fig:CRLB_step} (a) plots the the normalised CRLBs on delay estimation error in \eqref{eq:crlb_st_step} and \eqref{eq:crlb_stept} versus the received signal-to-noise ratio (SNR). Fig. \ref{fig:CRLB_step} (b) plots the normalised CRLBs on Doppler shift estimation error in \eqref{eq:crlb_sw_step} and \eqref{eq:crlb_stepw} versus the received SNR. The waveforms we focus on are the same as those for Fig. \ref{fig:AF_step} and Fig. \ref{fig:AF_step_psk}. As is illustrated in Fig. \ref{fig:CRLB_step} (a), the  CRLB on delay estimation error decreases by changing the phases of subpulses. The approximated CRLB for the un-modulated stepped frequency waveform in \eqref{eq:crlb_stept} is maximised since $f_0=0$ Hz and $\Delta f = 1/T$ Hz, thus any phase change will result in a lower delay CRLB, as is shown by \eqref{eq:crlb_st_step}. Agreeing with \eqref{eq:crlb_sw_step}, the CRLB on Doppler estimation error remains unchanged with phase modulation, as is illustrated in Fig. \ref{fig:CRLB_step} (b).

\section{Frequency Permutations and Phase Shift Keying Based Waveforms} \label{sec:pm_perm}
In this section, we introduce a novel frequency modulation on top of the phase modulation introduced in Section \ref{sec:pm_step} to embed more data into the waveform in \eqref{eq:signal_stepfreq}. In the linear stepped frequency radar waveform considered in \eqref{eq:signal_steppsk}, the frequency was linearly changed according to \eqref{eq:freqtones}. Motivated by \cite{Senanayake22TWC}, we proceed to change the order of frequency tones on top of the phase modulation, such that the randomisation in the frequency tones can be used to send more data. Interestingly, we observe that random permutation of the frequency tones results in a reasonably small AF side-lobe level in average and does not affect the radar local accuracy. 

\subsection{Incorporating Frequency Permutations} \label{sec:wf_freqperm_psk}
Consider the sequence of $\numfreqs$ frequency tones $f_0, f_1 , ..., f_{\numfreqs-1}$ illustrated by (\ref{eq:freqtones}). If a single frequency tone is used only once in a particular waveform having $\numfreqs$ subpulses, we can formulate $\numfreqs!$ of these waveforms. The frequencies of the $i$-th waveform, $\{f_0^i,...,f_{\numfreqs-1}^i \}$, follow the $i$-th permutation from a total of $\numfreqs!$ permutation sequences where $f_\indsubpulse ^i$ denotes the frequency in the $\indsubpulse $-th slot of the $i$-th waveform.
The $i$-th waveform can be expressed as
\begin{equation}
    s_i(t) = \sqrt{\frac{E}{\numfreqs T}}\sum_{\indsubpulse =0}^{\numfreqs-1}s_p(t-\indsubpulse T)\exp{(j(2\pi f^i_\indsubpulse  (t-\indsubpulse T)))}.  \label{eq:signal_lehmer}
\end{equation}
 To increase the achievable data rate, we introduce phase modulation which is performed independently of the frequency modulation. Note that we change the phase in a discrete manner from subpulse to subpulse, as discussed in Section \ref{sec:pm_step}, thus, allowing us to independently change the phase and frequency of each subpulse. As a result, if $\numphases$-ary phase shift keying is used, we can formulate $\numfreqs! \times \numphases^{\numfreqs}$ possible waveforms in total. The $i$-th waveform can now be expressed as:
\begin{equation}
    s_i(t) = \sqrt{\frac{E}{\numfreqs T}}\sum_{\indsubpulse =0}^{\numfreqs-1}s_p(t-\indsubpulse T)\exp{(j(2\pi f^i_\indsubpulse  (t-\indsubpulse T) + \theta_\indsubpulse ^i))},  \label{eq:signal_lehmerpsk}
\end{equation}
where $\theta_\indsubpulse ^i \in \{0, 2\pi/\numphases, ... , 2\pi(\numphases-1)/\numphases \}$ denotes the phase of the $\indsubpulse $-th subpulse of the $i$-th waveform and the frequencies of the $i$-th waveform follow one of the $\numfreqs!$ permutations given by
\begin{equation}
\begin{aligned}
\Psi_1 = \cdots = &\Psi_{\numphases^{\numfreqs}} = \{f_0,...,f_{\numfreqs-2},f_{\numfreqs-1}\}\\
\Psi_{\numphases^{\numfreqs}+1} = \cdots = &\Psi_{2\numphases^{\numfreqs}} = \{f_0, ..., f_{\numfreqs-1}, f_{\numfreqs-2}\}\\
&\vdots \\
\Psi_{(\numfreqs!-1) \times \numphases^{\numfreqs}+1} = \cdots = &\Psi_{\numfreqs! \times \numphases^{\numfreqs}} = \{f_{\numfreqs-1},...,f_{1},f_0\}.
\end{aligned}
\end{equation}
For simplicity we define $\numwaveforms = \numfreqs! \times \numphases^{\numfreqs}$ to be the total number of possible waveforms.

\subsection{Generation of the Waveforms}
As has been discussed, we can formulate $\numwaveforms$ possible waveforms under the novel modulation scheme in Section \ref{sec:wf_freqperm_psk}. When $\numfreqs$ and $\numphases$ are large, a huge look-up table is required to map the data to the waveform. However, we can avoid the use of the look-up table by implementing the mapping using the $\textit{Lehmer code}$ \cite{Knuth1973}. This was initially proposed in \cite{Senanayake22TWC} for the frequency permutation based waveforms but could be easily adapted for the case with both frequency and phase modulation.  To be more specific, if the incoming data symbol is represented as an index $i$ where $i \in \{1,..., \numwaveforms\}$, the integer $(\left \lfloor \frac{i-1}{\numphases^{\numfreqs}} \right \rfloor + 1)$ is mapped to a particular permutation of $\numfreqs$ frequency tones using a factorial number system where $\lfloor \cdot \rfloor$ denotes the operation of rounding the argument down. Then, the integer $(i - 1 - \left \lfloor \frac{i-1}{\numphases^{\numfreqs}}\right \rfloor \numphases^{\numfreqs})$ is converted to a sequence of $\numfreqs$ digits with base number $\numphases$ and sent into a standard \numphases PSK modulator to generate a particular sequence of $\numfreqs$ phases.  

\subsection{Ambiguity Function} \label{sec:AF_perm}
In order to derive the AF, we normalise the signal energy to one and re-express the $i$-th waveform in (\ref{eq:signal_lehmerpsk}) as
\begin{equation}
    s_i(t) = \sqrt{\frac{1}{\numfreqs T}}\sum_{\indsubpulse =0}^{\numfreqs-1}s_p(t-\indsubpulse T)\exp{(j(\omega^i_\indsubpulse  (t-\indsubpulse T) + \theta_\indsubpulse ^i))},  \label{eq:signal_lehmerpsk_radar}
\end{equation}
where $\omega^i_\indsubpulse  = 2\pi f_\indsubpulse ^i$ is the frequency in radians per second. Following similar steps as in Section \ref{sec:AF_step}, we can write the AF of the waveform in (\ref{eq:signal_lehmerpsk_radar}) as
\begin{equation}
A(\tau, \omega) = \left| \frac{1}{\numfreqs T}\sum_{\indsubpulse =0}^{\numfreqs-1}\sum_{n=0}^{\numfreqs-1}\hat{A}_p(\tau+(n-\indsubpulse )T, \omega - (\omega_n^i-\omega_\indsubpulse ^i)) e^{j(\omega \indsubpulse T+\omega_n^i(\tau+(n-\indsubpulse )T)+(\theta_\indsubpulse ^i-\theta_n^i))} \right|. \label{eq:AF_lehpsk}
\end{equation}
The zero-Doppler cut and the zero-delay cut can be expressed as
\begin{equation}
A(\tau, 0) = \left| \frac{1}{\numfreqs T} \sum_{\indsubpulse =0}^{\numfreqs-1}\sum_{n=0}^{\numfreqs-1}\hat{A}_p(\tau+(n-\indsubpulse )T, \omega_\indsubpulse ^i-\omega_n^i)  e^{j(\omega_n^i(\tau-(n-\indsubpulse )T)+(\theta_\indsubpulse ^i-\theta_n^i))} \right|, \label{eq:AF0dop_lehpsk}
\end{equation}
\begin{equation}
\begin{aligned}
A(0, \omega) = \left| \frac{1}{\numfreqs T} \sum_{\indsubpulse =0}^{\numfreqs-1}\sum_{n=0}^{\numfreqs-1}\hat{A}_p((n-\indsubpulse )T, \omega - (\omega_n^i-\omega_\indsubpulse ^i)) e^{j(\omega \indsubpulse T+\omega_n^i(n-\indsubpulse )T+(\theta_\indsubpulse ^i-\theta_n^i))} \right|.  \label{eq:AF0del_lehpsk}
\end{aligned}
\end{equation}
Similarly, the zero-delay cut in \eqref{eq:AF0del_lehpsk} can be simplified as
\begin{equation}
A(0,\omega) = \left| \frac{1}{\numfreqs T}\sum_{\indsubpulse =0}^{\numfreqs-1}\hat{A}_p(0, \omega)e^{j\omega \indsubpulse T} \right|, \label{eq:AF0delsimp_lehpsk}
\end{equation}
which implies that the zero-delay cut is not affected by frequency permutation and phase change.

As is analysed in \cite{Senanayake22TWC}, the overall structure of the AF resulting from the random stepped frequency permutation based waveform has a narrow main-lobe and small side-lobes. When phase modulation is introduced, these good radar properties will not be affected much. Such an AF can achieve good radar performance performance specially with clutter. Since transmitter tends to send a waveform with a randomly selected frequency permutation due to the randomness of incoming information, we can consider that the clutter enters via the average AF side-lobe level which is reasonably small for our waveform. The performance of the AFs with different parameters will be further discussed using numerical examples in Section \ref{sec:num_leh_radar}.

\subsection{Fisher Information Matrix and Cramer-Rao Lower Bounds} \label{sec:crlb_perm}
In this section we follow similar steps as is used in Section \ref{sec:crlb_step}. Replacing $\omega_m$ and $\theta_m$ in \eqref{eq:J11_step} to \eqref{eq:J22_step} with $\omega_m^i$ and $\theta_m^i$, respectively, the new FIM elements can be expressed as
\begin{equation}
    J_{11} \approx \frac{2CB}{\numfreqs T}\left( \numfreqs -\sum_{\indsubpulse =0}^{\numfreqs-2}\cos(\omega_0 T + \theta_\indsubpulse ^i-\theta_{\indsubpulse +1}^i) \right), \label{eq:J11}
\end{equation}
\begin{equation}
    J_{12} = J_{21} \approx -\frac{CT^2}{2}\sum_{\indsubpulse =0}^{\numfreqs-1}\omega_\indsubpulse ^i(2\indsubpulse +1), \label{eq:J12}
\end{equation}
\begin{equation}
    J_{22} = \frac{C\numfreqs^2T^2}{12}. \label{eq:J22}
\end{equation}

Based on \eqref{eq:J11}-\eqref{eq:J22}, the CRLBs of the delay estimation and Doppler shift estimation errors can be approximated by
\begin{equation}
\text{CRLB}_{\tau} \approx \frac{C^{-1}\numfreqs^2T}{2\numfreqs B\left(\numfreqs-\sum_{\indsubpulse =0}^{\numfreqs-2}\cos{(\omega_0 T+\theta_\indsubpulse ^i-\theta_{\indsubpulse +1}^i)}\right)-3T^3\left(\sum_{\indsubpulse =0}^{\numfreqs-1}(2\indsubpulse +1)\omega_\indsubpulse ^i \right)^2}, \label{eq:crlb_t}
\end{equation}
\begin{equation}
\text{CRLB}_{\omega} \approx \frac{24C^{-1}B\left(\numfreqs-\sum_{\indsubpulse =0}^{\numfreqs-2}\cos{(\omega_0 T+\theta_\indsubpulse ^i-\theta_{\indsubpulse +1}^i)}\right)}{2\numfreqs^2T^2B\left(\numfreqs-\sum_{\indsubpulse =0}^{\numfreqs-2}\cos{(\omega_0 T+\theta_\indsubpulse ^i-\theta_{\indsubpulse +1}^i)}\right)-3\numfreqs T^5\left(\sum_{\indsubpulse =0}^{\numfreqs-1}(2\indsubpulse +1)\omega_\indsubpulse ^i \right)^2}. \label{eq:crlb_w}
\end{equation}

By ignoring $J_{12}$ and $J_{21}$ \cite[eq.(10.94), (10.95)]{vantrees01}, \eqref{eq:crlb_t} and \eqref{eq:crlb_w} can be simplified to looser bounds, which can be expressed as
\begin{equation}
\text{CRLB}_{\tau} \approx \frac{C^{-1}\numfreqs T}{2B\left( \numfreqs -\sum_{\indsubpulse =0}^{\numfreqs-2}\cos(\omega_0 T + \theta_\indsubpulse ^i-\theta_{\indsubpulse +1}^i) \right)}, \label{eq:crlb_st}
\end{equation}
\begin{equation}
\text{CRLB}_{\omega} \approx \frac{12C^{-1}}{\numfreqs^2T^2}. \label{eq:crlb_sw}
\end{equation}
Again, the simplified expressions in (\ref{eq:crlb_st}) and (\ref{eq:crlb_sw}) clearly show the effect of parameters on the estimation errors. The delay estimation error lower bound in (\ref{eq:crlb_st}) is inversely proportional to B and to the effective bandwidth, $1/T$, of a single subpulse. The Doppler shift estimation error lower bound in (\ref{eq:crlb_sw}) is inversely proportional to the square of time duration of the whole waveform $\numfreqs T$. 
Compared to \eqref{eq:crlb_stept} and \eqref{eq:crlb_stepw}, the frequency permutation has no effect on both $\text{CRLB}_{\tau}$ and $\text{CRLB}_{\omega}$, which has been analysed in detail in \cite{Senanayake22TWC}. 
In addition, compared to \cite[eq.(48), (49)]{Senanayake22TWC}, it can be found that the phase modulation has exactly no effect on the simplified $\text{CRLB}_{\omega}$ while it does affect $\text{CRLB}_{\tau}$. Similar to Section \ref{sec:crlb_step}, the maximum value and minimum value of $\text{CRLB}_{\tau}$ in \eqref{eq:crlb_sw} are kept unchanged when phase modulation is introduced. The analysis is supported by the numerical examples in Section \ref{sec:num_leh_radar}. 

\subsection{Numerical Examples}\label{sec:num_leh_radar}
\begin{figure*}
    \centerline{\includegraphics[width=17.3cm,height=6.7cm]{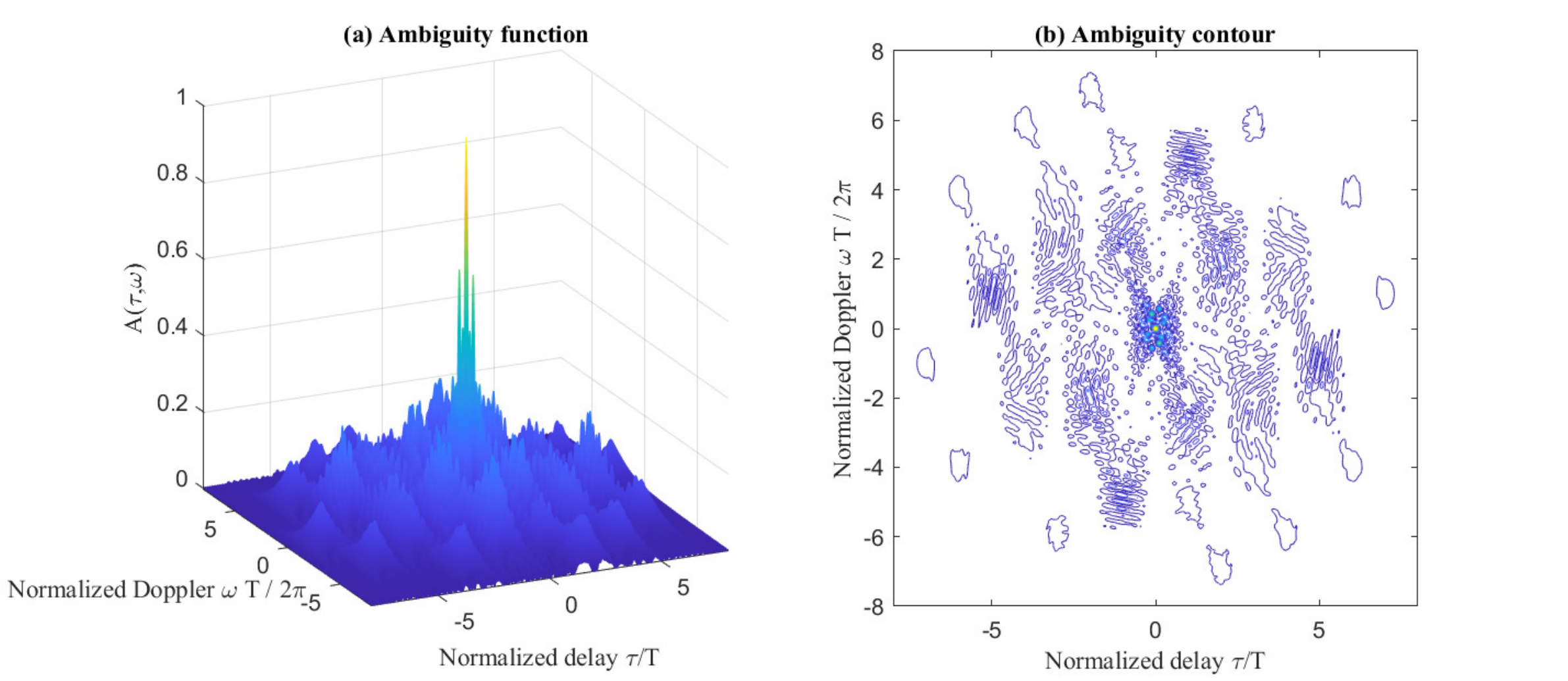}}
    \caption{AF of a frequency permutation based waveform for $\numfreqs = 8$. The frequency sequence is $[f_2,f_7,f_5,f_0,f_4,f_1,f_6,f_3]$.
    }\label{fig:AF_perm}
\end{figure*}

\begin{figure*}
    \centerline{\includegraphics[width=17.3cm,height=6.7cm]{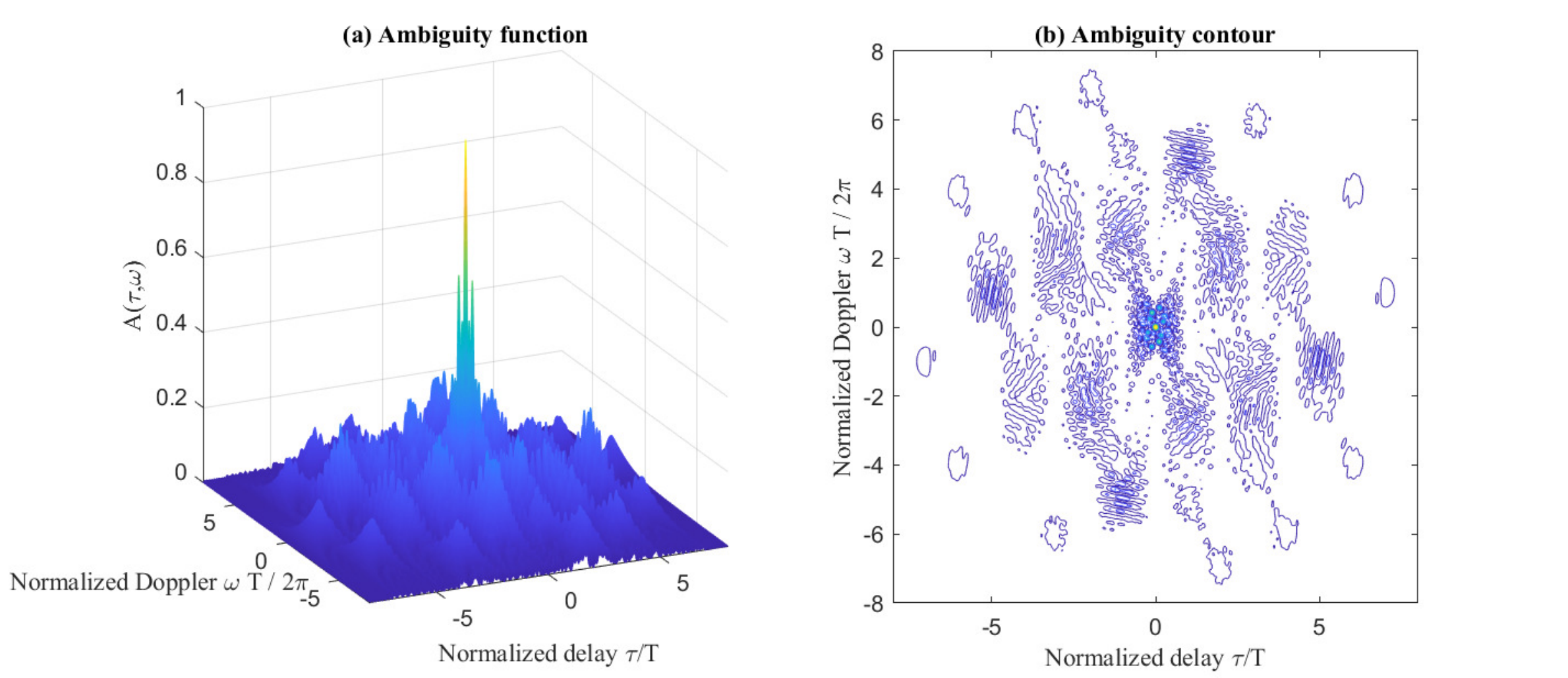}}
    \caption{AF of a frequency permutation and PSK based waveform for $\numfreqs = 8$ and $\numphases=4$. The frequency sequence is $[f_2,f_7,f_5,f_0,f_4,f_1,f_6,f_3]$. The phase sequence is $[3\pi/4,3\pi/4,3\pi/4,\pi/2,3\pi/4,\pi/2,\pi/4,\pi/2]$.
    }\label{fig:AF_perm_psk}
\end{figure*}

In this section, we provide numerical examples to verify the theoretical analyses in Section \ref{sec:AF_perm} and Section \ref{sec:crlb_perm}. Fig. \ref{fig:AF_perm} and Fig. \ref{fig:AF_perm_psk} illustrate the AFs and the corresponding contour plots for the random frequency permutation based waveforms for $\numfreqs = 8$ without phase modulation and with phase modulation, respectively. 
The phase sequence of the waveform in Fig. \ref{fig:AF_perm_psk} is the same as that in Fig. \ref{fig:AF_step_psk} and the frequency permutation is randomly generated. Both of them will be fixed in the AF and CRLB examples for the remainder of this section.
We observe that the main-lobes of the AFs in Fig. \ref{fig:AF_perm} and Fig. \ref{fig:AF_perm_psk} are very similar, which again implies that the phase modulation has negligible impact on the AF around the origin. In addition, by comparing Fig. \ref{fig:AF_perm} and Fig. \ref{fig:AF_perm_psk} with Fig. \ref{fig:AF_step} and Fig. \ref{fig:AF_step_psk}, it can be observed that the frequency permutation has a larger impact on the broader structure of the AF than the phase modulation does. By randomly choosing a frequency permutation, the volume under the AF is more likely to be evenly spread over the delay-Doppler plane as is evident by the side-lobe structure of Fig. \ref{fig:AF_perm} and Fig. \ref{fig:AF_perm_psk} compared to that of a waveform with an ascending frequency sequence given in Fig. \ref{fig:AF_step} and Fig. \ref{fig:AF_step_psk}. This result shows the averaging effect on the AF side-lobe level discussed in Section \ref{sec:AF_perm}, which supports the statement that the performance of this waveform with clutter is good on average. 

\begin{figure*}
    \centerline{\includegraphics[width=19.5cm,height=7cm]{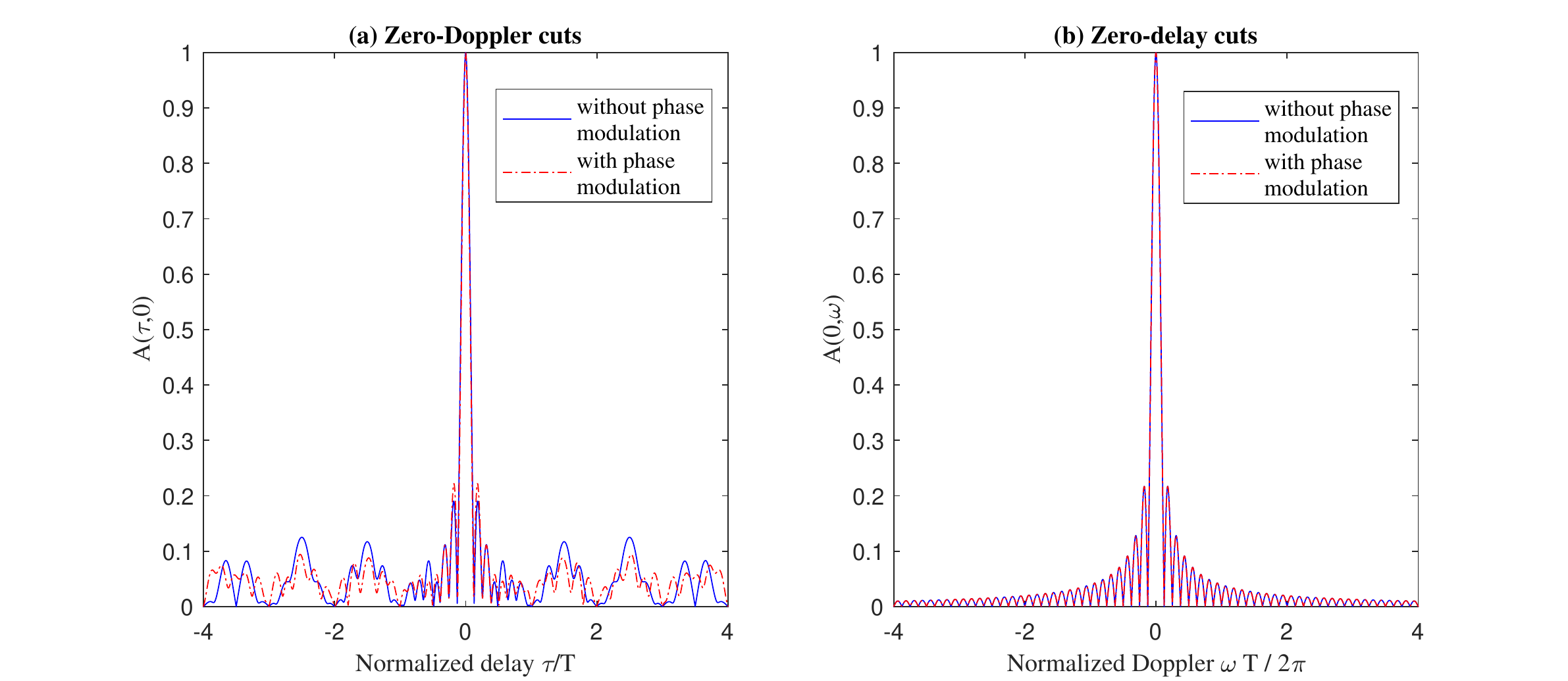}}
    \caption{The (a) zero-Doppler cuts and (b) zero-delay cuts of the AFs in Fig. \ref{fig:AF_perm} and Fig. \ref{fig:AF_perm_psk}. 
    }\label{fig:0cuts_perm}
\end{figure*}

Fig. \ref{fig:0cuts_perm} plots the one dimensional cuts of the AFs in Fig. \ref{fig:AF_perm} and Fig. \ref{fig:AF_perm_psk}. The zero-Doppler cuts in Fig. \ref{fig:0cuts_perm} (a) indicate that the change due to the phase modulation in both the highest side-lobe level and  the curvature at the origin  is negligible. As is discussed in Section \ref{sec:num_step_radar}, the zero-delay cut is not affected by the phase modulation, which can also be observed in Fig. \ref{fig:0cuts_perm} (b). Compared to Fig. \ref{fig:0cuts_step}, we notice that the frequency permutation does not affect the zero-delay cuts as well. This agrees with the expression in \eqref{eq:AF0delsimp_lehpsk} as well as the analysis in \eqref{eq:crlb_sw}.

\begin{figure*}
    \centerline{\includegraphics[width=17.3cm,height=6.7cm]{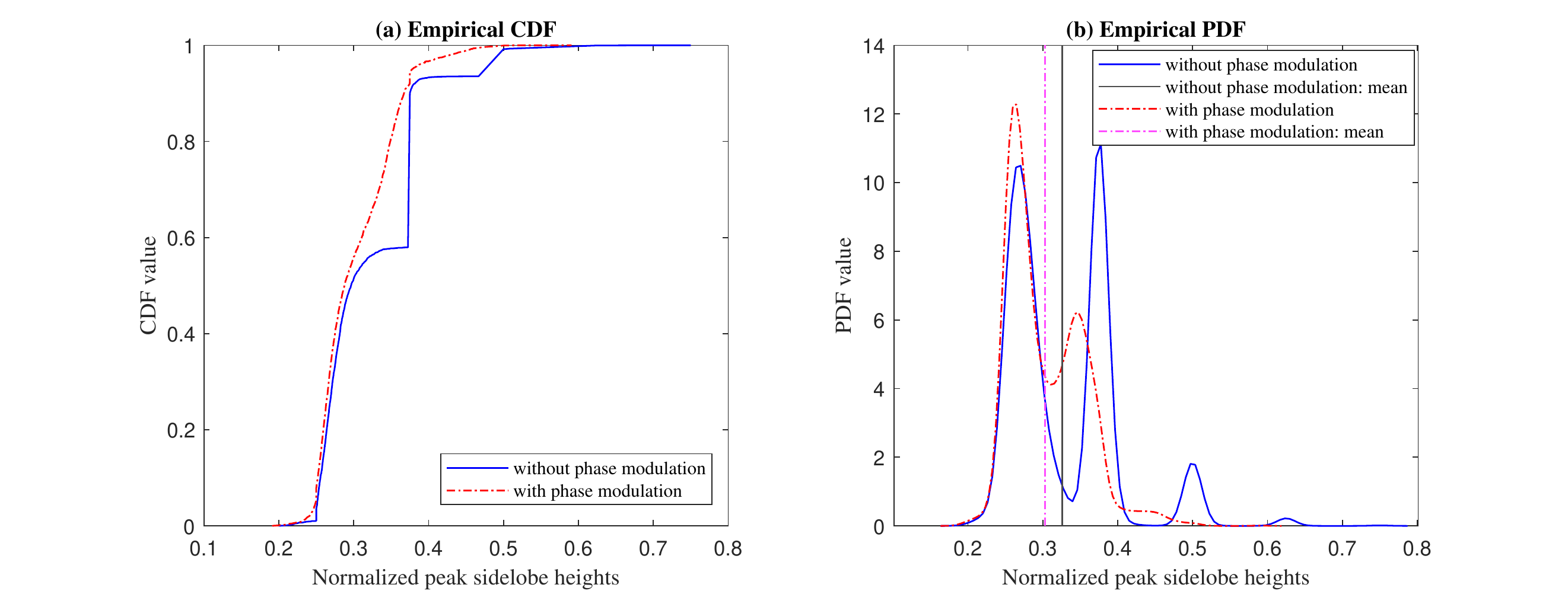}}
    \caption{The (a) empirical CDF and (b) empirical PDF of the normalised peak side-lobe levels of AFs with and without phase modulation for $\numfreqs = 8$ and $\numphases = 4$. 
    }\label{fig:AF_sielobdes}
\end{figure*}

Fig. \ref{fig:AF_perm}, Fig. \ref{fig:AF_perm_psk} and Fig. \ref{fig:0cuts_perm} illustrate the comparison of side-lobe levels using two example AFs. In order to analyse the impact of phase modulation on the side-lobe performances of all possible waveforms, Fig. \ref{fig:AF_sielobdes} plots the empirical cumulative distribution function (CDF) and empirical probability density function (PDF) of the normalised peak side-lobe levels of the AFs with and without phase modulation for $\numfreqs = 8$ and $\numphases = 4$. When phase modulation is introduced, both the CDF and PDF become smoother. Besides, from Fig. \ref{fig:AF_sielobdes} (b), we observe that the empirical mean of the peak sidelobe level is slightly decreased from $0.326$ to $0.303$ with phase modulation. This clearly illustrates that phase modulation does not have significant effect on the average side-lobe levels of the AFs.

\begin{figure*} 
    \centerline{\includegraphics[width=19.5cm,height=7.5cm]{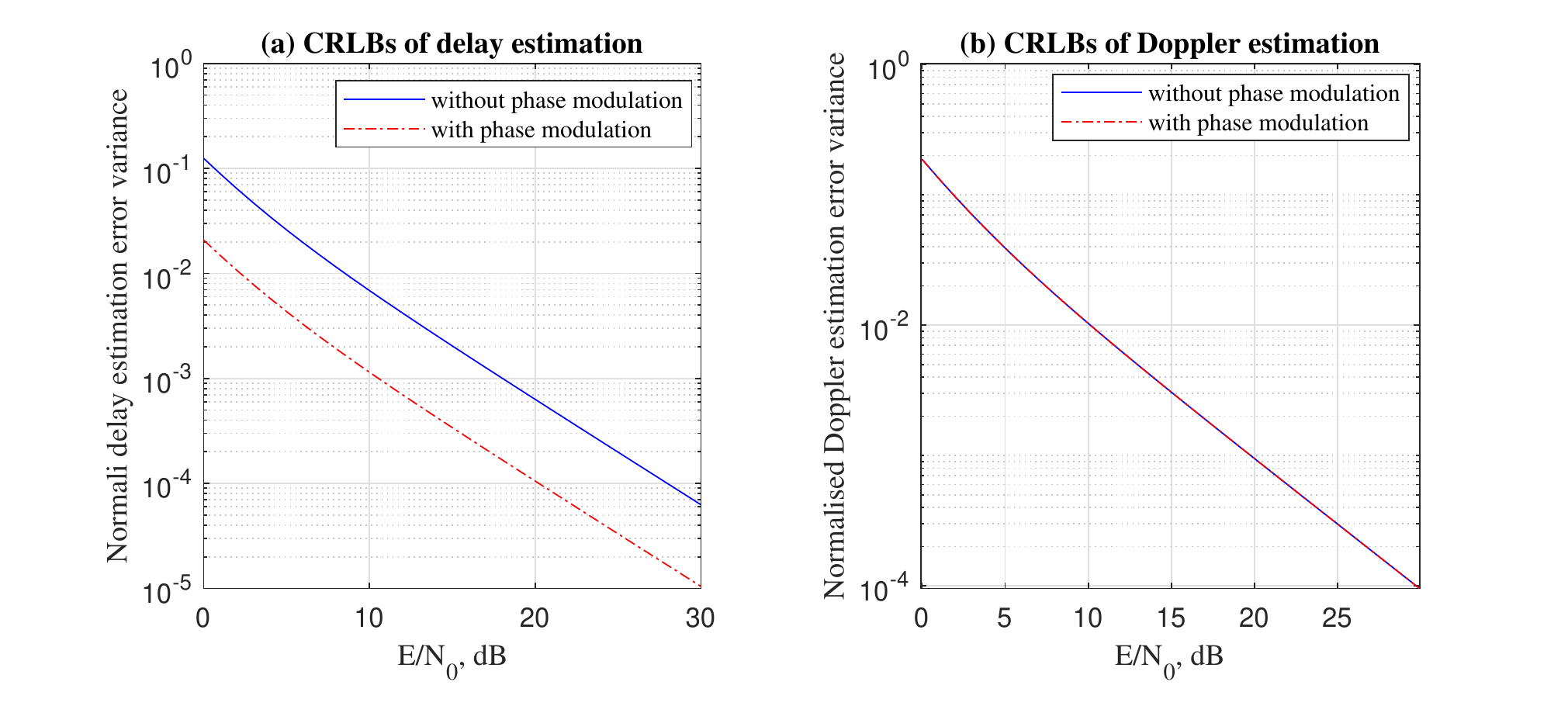}}
    \caption{{The (a) approximated CRLBs in \eqref{eq:crlb_st} on delay estimation errors and (b) approximated CRLBs in \eqref{eq:crlb_sw} on Doppler shift estimation errors for stepped frequency based waveforms with and without phase modulation.}. 
    }\label{fig:CRLB_perm}
\end{figure*}

Fig. \ref{fig:CRLB_perm} plots the the normalised CRLBs on delay estimation error in \cite[eq.(48)]{Senanayake22TWC} and \eqref{eq:crlb_st}, as well as the normalised CRLBs on Doppler shift estimation error in \cite[eq.(49)]{Senanayake22TWC} and \eqref{eq:crlb_sw} versus the received SNR, respectively. The waveforms we focus on are the same as those for Fig. \ref{fig:AF_perm} and Fig. \ref{fig:AF_perm_psk}. Since $f_0=0$ Hz and $\Delta f = 1/T$ Hz, the simplified $\text{CRLB}_{\tau}$ in \eqref{eq:crlb_st} is maximised without phase modulation and is decreased due to the change of phases, which is shown in Fig. \ref{fig:CRLB_perm} (a). The simplified $\text{CRLB}_{\omega}$ in \eqref{eq:crlb_sw} remains unchanged with phase change, as is shown in Fig. \ref{fig:CRLB_perm} (b).

\section{Communications Performance Analysis} \label{sec:comms_performance}
In this section, we focus on the communications performance of the proposed waveform. In Section \ref{sec:pm_step}, we consider the traditional MPSK modulation for which the communication performance is quite well known and therefore is not discussed in this paper. In Section \ref{sec:pm_perm} we consider combining MPSK modulation with frequency permutation, which results in a new signalling scheme. In this section we focus on this interesting new communication system and analyse its performance in terms of the block error probability. We perform the optimal ML detection at the communication receiver and derive analytical expressions for the union bound, nearest neighbour approximation and a new upper bound on the block error probability. In addition, we propose the implementation of an efficient communication receiver based on the Hungarian algorithm.

While we only focus on the block error probability performance, we should notice that the achievable data rate of the proposed waveform depends on the total number of waveforms we can generate which is  $\lfloor \log_2(\numwaveforms) \rfloor / \numfreqs T$ bits/second. This value is much higher compared to the number of waveforms that can be generated under the frequency permutation method in \cite{Senanayake22TWC}.

\subsection{Maximum Likelihood Detection}
Assume there are $N$ antennas at the communications receiver. The baseband received signal at time $t$ can be represented by an $N \times 1$ vector
\begin{equation}
\boldsymbol{r}_c(t) = \boldsymbol{h} s_i(t) + \boldsymbol{n}(t),
\end{equation}
where $\boldsymbol{h}$ is the fading channel vector, $s_i(t)$ is the transmitted signal and $\boldsymbol{n}(t)$ is an AWGN vector in which the power spectral density (PSD) of each element is $N_0$. The transmission of $s_i(t)$ is assumed to be equally likely among all $\numwaveforms$ possible waveforms. Suppose the channel vector is known at the receiver, then ML detection of the received signal is given by
\begin{equation}
\begin{aligned}
\hat{s}_i(t) &= \argmin_{s_k(t)\in \{s_1(t),...,s_{\numwaveforms}(t) \}} \int_{0}^{\numfreqs T} |\boldsymbol{h}^H\boldsymbol{r}_c(t) - \boldsymbol{h}^H\boldsymbol{h}s_k(t)|^2 dt\\
&= \argmax_{s_k(t)\in \{s_1(t),...,s_{\numwaveforms}(t)\}} \Re\left\{\int_{0}^{\numfreqs T} s_k^*(t)\boldsymbol{h}^H\boldsymbol{r}_c(t) dt \right\}, \label{eq:ML}
\end{aligned}
\end{equation}
where $\Re\{\cdot \}$ denotes the operation of taking the real part of the argument.

\subsection{Error Probability Analysis}  \label{sec:pe}
\subsubsection{AWGN channel} \label{sec:pe_awgn}
The AWGN channel is one of the fundamental channel models. By assuming a unit channel gain, the fading vector satisfies $\boldsymbol{h}^H\boldsymbol{h}=N$.
Suppose the $i$-th waveform, $s_i(t)$, is transmitted. Then the probability of the correct detection of $s_i(t)$ is given by
\begin{equation}
P_c(i) = \Pr \left[\xi_{ii} = \max_{k\in\{1,...,\numwaveforms \}}  \xi_{ik}\right],
\end{equation}
where $\xi_{ik}$ is the $k$-th decision variable when $s_i(t)$ is transmitted, and is given by
\begin{equation}
\xi_{ik} = \Re\left\{\int_{0}^{\numfreqs T} s_k^*(t)\boldsymbol{h}^H\boldsymbol{r}_c(t) dt\right\}. \label{eq:dv}
\end{equation}
Since the transmission of each waveform is assumed to be equally likely, the average probability of making a wrong detection is 
\begin{equation}
P_e = \frac{1}{\numwaveforms} \sum_{i=1}^{\numwaveforms} \left(1- P_c(i) \right).
\end{equation}
We note that it is difficult to calculate the exact expression of $P_e$ since it requires multi-dimensional integrals. Therefore, we consider the union bound on $P_e$, which is given by
\begin{equation}
    P_e \leq P_e^{UB} = \frac{1}{\numwaveforms} \sum_{i=1}^{\numwaveforms} \sum_{\substack{k=1, \\ k\neq i}}^{\numwaveforms} P_{ik},
\end{equation}
where $P_{ik}$ denotes the pairwise error probability (PEP) that the $k$-th signal is preferred over the $i$-th signal when the $i$-th signal is transmitted and is given by
\begin{equation}
P_{ik} = \Pr \left[\xi_{ik} - \xi_{ii} \geq 0\right]. \label{eq:pik1}
\end{equation}
Due to the symmetry of PEPs, we can further simplify the union bound as
\begin{equation}
P_e^{UB} = \sum_{k=2}^{\numwaveforms} P_{1k}. \label{eq:UB}
\end{equation}

Similar to \cite[eq.(10)-(11)]{Senanayake22TWC}, next we proceed to analyse $P_{ik}$, by substituting \eqref{eq:dv} into \eqref{eq:pik1} and rearranging the expressions to obtain
\begin{equation}
P_{ik} = \Pr \left[ \frac{E\left(\numfreqs-\sum_{q=1}^{\numfreqs-d} \cos(\theta_{\indsubpulse (q)}^k-\theta_{\indsubpulse (q)}^i) \right)}{\numfreqs}\boldsymbol{h}^H\boldsymbol{h} < Z_1\right],
\end{equation}
where $d$ is the number of frequency tones $s_i(t)$ and $s_k(t)$ differ by, $\indsubpulse (q)$ is the index of the $q$-th subpulse where $s_i(t)$ and $s_k(t)$ have the same frequency and $Z_1 = \Re\left\{ \boldsymbol{h}^H\int_{0}^{\numfreqs T}(s_k^*(t)-s_i^*(t))\boldsymbol{n}(t)dt \right\}$. When conditioned on the channel fading, $Z_1$ is a Gaussian random variable with zero mean and variance $E\left(\numfreqs-\sum_{q=1}^{\numfreqs-d} \cos(\theta_{\indsubpulse (q)}^k-\theta_{\indsubpulse (q)}^i) \right)N_0\boldsymbol{h}^H\boldsymbol{h} / \numfreqs$. Thus we can express the PEP in \eqref{eq:pik1} as
\begin{equation}
P_{ik} = \Pr \left[\sqrt{\boldsymbol{h}^H\boldsymbol{h}} \leq  \alpha_{ik}Z\right], \label{eq:pik}
\end{equation}
where  
\begin{equation*}
\alpha_{ik} = \sqrt{\frac{\numfreqs N_0}{E\left(\numfreqs-\sum_{q=1}^{\numfreqs-d} \cos(\theta_{\indsubpulse (q)}^k-\theta_{\indsubpulse (q)}^i) \right)}},
\end{equation*}
and $Z \sim \mathcal{N}(0,1)$ is a standard Gaussian random variable. By setting $\boldsymbol{h}^H\boldsymbol{h}=N$, the pairwise error probability $P_{ik}$ over the AWGN channel can be represented as
\begin{equation}
P_{ik} = Q\left( \frac{\sqrt{N}}{\alpha_{ik}} \right), \label{eq:pik_awgn}
\end{equation}
where $Q(\cdot)$ is the Gaussian Q-function.
Substituting \eqref{eq:pik_awgn} to \eqref{eq:UB} we find the resulting union bound over the AWGN channel as
\begin{equation}
P_e^{UB} = \sum_{k=2}^{\numwaveforms} Q\left( \frac{\sqrt{N}}{\alpha_{1k}} \right). \label{eq:ub_awgn}
\end{equation}

We note that, if $\numfreqs$ and $\numphases$ are large, there are $\numwaveforms -1$ pairwise error probabilities to be added up, which will make the union bound far above the exact error probability, especially in the low SNR region. To provide a better approximation, we derive the nearest neighbour approximation of the error probability, in which we only consider the pairwise error probabilities corresponding to the nearest neighbours \cite{Proakis08}. When $\numphases > 2$, the nearest neighbours of a particular waveform $s_i(t)$ are the waveforms with the same frequency permutation as $s_i(t)$, while the phase is only different in a single subpulse by the amount of $2\pi/\numphases$. The number of nearest neighbours for a particular waveform is $2\numfreqs$. Thus the nearest neighbour approximation can be expressed as
\begin{equation}
P_e^{NN} = 2\numfreqs P_{1k}^{NN}, \label{eq:NN}
\end{equation}
where $P_{1k}^{NN}$ is the PEP that the nearest neighbour $s_k(t)$ is preferred over $s_1(t)$ when $s_1(t)$ is transmitted, which can be expressed as
\begin{equation}
P_{1k}^{NN} = Q\left( \frac{\sqrt{N}}{\alpha_{ik}^{NN}} \right), \label{eq:pikNN_awgn}
\end{equation}
where 
\begin{equation}
\alpha_{ik}^{NN} = \sqrt{\frac{\numfreqs N_0}{E\left(1-\cos(\frac{2\pi}{\numphases}) \right)}}. \label{eq:alphaNN}
\end{equation}
When $\numphases = 2$, however, the set of nearest neighbours of a particular waveform $s_i(t)$ is the union of two subsets. The first subset has a size of $\numfreqs$, which contains the waveforms with the same frequency permutation as $s_i(t)$, while the phase is only different in a single subpulse by the amount of $\pi$. In the second subset, the waveforms have $\numfreqs - 2$ subpulses in the same positions as $s_i(t)$. For the other two subpulses, the order of the frequency tones are different compared to $s_i(t)$ while phases do not matter. The size of the second subset is $2\numfreqs(\numfreqs-1)$. Therefore, when $\numphases = 2$, the nearest neighbour approximation can be expressed as
\begin{equation}
P_e^{NN} = (2\numfreqs^2-\numfreqs) P_{1k}^{NN}, \label{eq:NN_Mtilde2}
\end{equation}
where $P_{1k}^{NN}$ in \eqref{eq:NN_Mtilde2} can be expressed by \eqref{eq:pikNN_awgn} with
$\alpha_{ik}^{NN} = \sqrt{\frac{\numfreqs N_0}{2E}}$. 
Thus, the nearest neighbour approximation over the AWGN channel can be written as
\begin{equation}
\begin{aligned}
P_e^{NN} = \left\{\begin{aligned}
&2\numfreqs Q\left( \sqrt{\frac{NE\left(1-\cos(2\pi/\numphases)\right)}{LN_0}} \right), & \quad \numphases > 2\\
&(2\numfreqs^2-\numfreqs) Q\left( \sqrt{\frac{2NE}{LN_0}} \right), & \quad \numphases = 2.   
\end{aligned}
\right.  \label{eq:nn_awgn}
\end{aligned}
\end{equation}


\subsubsection{Correlated fading channels}  \label{sec:pe_fading}
 
Under fading channels, we first focus on the correlated Rician fading channel with a line-of-sight (LOS) path and a scattered path. For such a system, the channel fading vector can be expressed as
\begin{equation}
\boldsymbol{h} = \sqrt{\frac{K}{K+1}}\Delta + \sqrt{\frac{1}{K+1}}\boldsymbol{R}_u^{1/2} \boldsymbol{u}, \label{eq:rician_channel_vector}
\end{equation}
where $\Delta$ denotes the $N\times1$ complex LOS phase vector with the $i$-th element having the property $|\Delta_i|^2 = 1$, $\boldsymbol{u} \sim \mathcal{CN}(0,\boldsymbol{I}_N)$, $K$ is the Rician factor which denotes the relative strength of the LOS path to the scattered path and $\boldsymbol{R}_u$ denotes the $N\times N$ correlation matrix of the scattered component. As  $\boldsymbol{R}_u$ is a symmetric matrix it can be expressed as
\begin{equation}
\boldsymbol{R}_u = \boldsymbol{V \Omega V}^H,  \label{eq:eig_decomp_Ru}
\end{equation}
where $\boldsymbol{V}$ is unitary and $\boldsymbol{\Omega}=\text{diag}(\lambda_1, ..., \lambda_N)$ is a diagonal matrix containing the eigenvalues $\lambda_1, ..., \lambda_N$ of $\boldsymbol{R}_u$. We denote the $j$-th element of the vector $(\boldsymbol{V}^H \Delta)$ as $(\boldsymbol{V}^H \Delta)_j$. Whilst not given here due to page limitations, after lengthy mathematical manipulations,  $P_{1k}$ for correlated Rician channels can be derived as
\begin{equation}
\begin{aligned}
P_{1k} = \frac{1}{\pi} &\int_0^{\pi/2} \prod_{j=1}^N  \left( \frac{2(K+1) \alpha_{1k}^2 sin^2\theta}{\lambda_j + 2(K+1) \alpha_{1k}^2 sin^2\theta}\right)  \exp \left( -K\sum_{j=1}^N \frac{|(\boldsymbol{V}^H \Delta)_j|^2}{\lambda_j + 2(K+1) \alpha_{1k}^2 sin^2\theta} \right) d\theta.
\label{eq:Ric_pep}
\end{aligned}
\end{equation}
and the full proof of the above derivation is included in Appendix \ref{sec:ricianpep}.
Substituting \eqref{eq:Ric_pep} into \eqref{eq:UB}, \eqref{eq:NN} and \eqref{eq:NN_Mtilde2}, we obtain the union bound as well as the nearest neighbour approximation of the error probability over a correlated Rician fading channel.


Next, we focus on the Rayleigh fading channel, which is a special case of the Rician fading channel. The fading vector $\boldsymbol{h}$ for the Rayleigh channel does not have the LOS component, and is given by \eqref{eq:rician_channel_vector} with $K=0$. Therefore, by setting $K=0$ in \eqref{eq:Ric_pep}, the pairwise error probability can be expressed as
\begin{equation}
P_{1k} = \frac{1}{\pi} \int_0^{\pi/2} \prod_{j=1}^N  \left( \frac{2 \alpha_{1k}^2 sin^2\theta}{\lambda_j + 2 \alpha_{1k}^2 sin^2\theta}\right) d\theta.
\label{eq:Ray_pep}
\end{equation}

As has been stated in Section \ref{sec:pe_awgn}, when $\numfreqs$ and $\numphases$ are large, the union bound can be very loose, especially in the low SNR regime. This problem becomes more severe in fading channels, since the channels are random vectors. Therefore, in the following we propose a tighter upper bound on the error probability for fading channels as 
\begin{equation}
    P_e \leq P_e^{NB} = \min_{\gamma \geq 0} \left\{\Pr \left[\boldsymbol{h}^H\boldsymbol{h}<\gamma \right] + \Pr \left[\boldsymbol{h}^H\boldsymbol{h}\geq \gamma \right]\sum_{k=2}^{\numwaveforms} \Tilde{P}_{1k} \right\}, \label{eq:newUB}
\end{equation}
where $\gamma$ is a threshold defined for the total channel gain and 
\begin{equation}
\Tilde{P}_{1k} = \Pr\left[\xi_{11}<\xi_{1k} \mid \boldsymbol{h}^H\boldsymbol{h}\geq \gamma \right]. \label{eq:conditioned_pik}
\end{equation}
In \eqref{eq:newUB}, the bound is created by assuming that a detection error always occurs when the channel gain is lower than a defined threshold $\gamma$. The minimisation with respect to $\gamma$ in \eqref{eq:newUB} is taken to make the upper bound as tight as possible. Using the result in \eqref{eq:pik} and following the steps in Appendix \ref{sec:conditioned_pik}, \eqref{eq:conditioned_pik} can be expressed as
\begin{equation}
\Tilde{P}_{1k} = \frac{1}{\Pr[\boldsymbol{h}^H\boldsymbol{h} \geq \gamma]} \int_{\gamma}^{\infty} \frac{1}{\pi} \int_{0}^{\pi/2} e^{-\frac{x}{2\alpha_{1k}^2sin^2\theta}} d\theta f_{\boldsymbol{h}^H\boldsymbol{h}}(x) dx, \label{eq:conditioned_pik1}
\end{equation}
where $f_{\boldsymbol{h}^H\boldsymbol{h}}(x)$ is the probability density function of $\boldsymbol{h}^H\boldsymbol{h}$. 
   
For the correlated Rayleigh fading channel, we can rewrite $\boldsymbol{h}^H\boldsymbol{h}$ as
\begin{equation}
    \boldsymbol{h}^H\boldsymbol{h} = \boldsymbol{u}^H \boldsymbol{R}_u \boldsymbol{u} \equiv \sum_{j=1}^N \lambda_j |\boldsymbol{u}_j|^2,  \label{eq:Ray_hHh}
\end{equation}
where $\boldsymbol{u}_j$ is the $j$-th element of the vector $\boldsymbol{u}$ and $\equiv$ indicates statistical equivalence. Thus, the probability density function of the random variable in \eqref{eq:Ray_hHh} can be expressed as \cite{johnson1970}
\begin{equation}
\begin{aligned}
f^{\text{Ray}}_{\boldsymbol{h}^H\boldsymbol{h}}(x) = \left\{\begin{aligned}
&\sum_{j=1}^N \frac{b_j}{\lambda_j} e^{-x/\lambda_j}, &\quad x \geq 0\\
&0, &\quad x < 0,
\end{aligned}
\right.  
\end{aligned} \label{eq:pdf_ray_hHh}
\end{equation}
where $b_j = \lambda_j^{N-1} \prod_{\substack{n=1 \\ n \neq j}}^N 1/(\lambda_j-\lambda_n)$. Hence the cumulative distribution function of $\boldsymbol{h}^H\boldsymbol{h}$ is given by
\begin{equation}
\begin{aligned}
F^{\text{Ray}}_{\boldsymbol{h}^H\boldsymbol{h}}(x) = \left\{\begin{aligned}
&\sum_{j=1}^N b_j (1-e^{-x/\lambda_j}), &\quad x \geq 0\\
&0, &\quad x < 0.
\end{aligned}
\right.  
\end{aligned} \label{eq:cdf_ray_hHh}
\end{equation}
Substituting \eqref{eq:pdf_ray_hHh} and \eqref{eq:cdf_ray_hHh} into \eqref{eq:conditioned_pik1} and \eqref{eq:newUB}, we can express the new upper bound over the correlated Rayleigh fading channel as 
\begin{equation}
    P_e^{NB} = \min_{\gamma \geq 0} \left\{ \sum_{j=1}^N b_j (1-e^{-x/\lambda_j}) +\frac{1}{\pi} \sum_{k=2}^{\numwaveforms}\sum_{j=1}^N \frac{2b_j\alpha_{1k}sin^2\theta}{\lambda_j + 2\alpha_{1k}sin^2\theta} \int_0^{\pi/2}  e^{-\gamma\left( \frac{1}{2\alpha_{1k}sin^2\theta} + \frac{1}{\lambda_j}\right)} d\theta\right\}. \label{eq:Ray_NB}
\end{equation}

\subsection{Optimal Communications Receiver Implementation}
In this section we propose the implementation of an efficient communications receiver for the novel signalling scheme presented in Section \ref{sec:pm_perm}. The optimal receiver based on the ML detection rule in (\ref{eq:ML}) looks for the maximum output over the correlations between the received signal and all possible transmitted signals. If an exhaustive search is applied to find the maximum, the receiver has a worst case complexity of $\mathcal{O}(\numwaveforms)$, which means the detection process will be prohibitively slow when $\numfreqs$ and $\numphases$ are large. Instead of correlating the whole received signal with the reference signals as in \eqref{eq:ML}, an efficient receiver can be implemented using the correlation in each subpulse
\begin{equation}
x_{\Bar{m},\indsubpulse} = \Re \left \{\int_{(\indsubpulse-1)T}^{\indsubpulse T} \boldsymbol{h}^H\boldsymbol{r}_c(t) \phi_{\Bar{m}}(t-(\indsubpulse -1)T)dt \right \}, \label{eq:R_entry}
\end{equation}
to formulate the matrix
\begin{equation}
\boldsymbol{X} = \left(x_{\Bar{m},\indsubpulse } \right) \in \mathbb{R}^{(\numphases \numfreqs) \times \numfreqs}. \label{eq:R}
\end{equation}
The basis function $\phi_{\Bar{m}}(t)$ is defined as
\begin{equation*}
\phi_{\Bar{m}}(t) = s_p(t)\exp(j(2\pi f_{n-1} t + \theta_{m})),
\end{equation*}
where $\Bar{m} \in \{1,2,...,\numphases \numfreqs\}$ is the row index of the $(\Bar{m},l)$-th element in $\boldsymbol{X}$, $\indsubpulse \in \{1,2,...,\numfreqs\}$ is the column index of the $(\Bar{m},l)$-th element in $\boldsymbol{X}$ which denotes the index of the subpulse, $n=\left(\lfloor \frac{\Bar{m}-1}{\numphases } \rfloor + 1\right) \in \{1,2,...,\numfreqs\}$ denotes the index of the frequency of the basis function $\phi_{\Bar{m}}(t)$, $m=\left(\Bar{m}-\lfloor \frac{\Bar{m}-1}{\numphases } \rfloor \numphases\right) \in \{1,2,...,\numphases \}$ and $\theta_{m} = 2 \pi (m-1) / \numphases$ is the phase of  $\phi_{\Bar{m}}(t)$.

Then, we split $\boldsymbol{X}$ into $\numfreqs^2$ blocks of $\numphases $ elements whose basis functions have the same frequency and time slot. Using the maximum element in each block, we can formulate a new matrix, which can be expressed as
\begin{equation}
\boldsymbol{Y} = (y_{n,\indsubpulse }) \in \mathbb{R}^{\numfreqs \times \numfreqs},
\end{equation}
where
\begin{equation}
y_{n,\indsubpulse } = \max_{\Bar{m} \in \{\numphases (n-1)+1, ..., \numphases n\}} x_{\Bar{m},\indsubpulse }.
\end{equation}
The row indices of $y_{n,\indsubpulse }$ in the original matrix $\boldsymbol{X}$ is stored separately in $\Bar{m}_{n,\indsubpulse }$. Note that the basis function used for $y_{n,\indsubpulse }$ is a complex exponential with frequency $f_{n-1}$. The worst case complexity of finding $\numfreqs^2$ maximums in $\numfreqs^2$ blocks is $\mathcal{O}(\numfreqs^2\times \numphases )$ if an exhaustive search is used. 

To detect the frequency of each subpulse, we apply the Hungarian algorithm to $\boldsymbol{Y}$ to select $\numfreqs$ elements $\hat{r}_{\indsubpulse }$, $\indsubpulse \in \{1,...,L\}$, such that these elements are in $\numfreqs$ different rows and $\numfreqs$ different columns, and the sum of the $\numfreqs$ elements is maximised. The detection of the frequency $\hat{f}^i_{\indsubpulse -1}$ is determined by the row index $\hat{n}_\indsubpulse $ of the selected element in the $\indsubpulse $-th column. Note that the Hungarian algorithm has a worst case time complexity of $\mathcal{O}(\numfreqs^3)$ \cite{kuhn55}.

To detect the phase of each subpulse, we use the recorded indices $\Bar{m}_{n,\indsubpulse }$ of the $\numfreqs$ elements selected by the Hungarian method. The detection of the phase $\hat{\theta}^i_{\indsubpulse -1}$ is determined by the row index of the selected element in the $\indsubpulse $-th column of the original matrix $\boldsymbol{X}$.

Algorithm \ref{algo:hun} summarises the main steps of the Hungarian algorithm based communication receiver. The proposed optimal receiver is an efficient implementation of the optimal ML detector.

\begin{algorithm}[t]
 \caption{Hungarian Algorithm Based Method for the Communication Receiver} \label{algo:hun}
 \begin{algorithmic}[1]
  \STATE Formulate $\boldsymbol{X} = \left(x_{\Bar{m},\indsubpulse } \right) \in \mathbb{R}^{(\numphases \numfreqs) \times \numfreqs}$ using \eqref{eq:R_entry} and \eqref{eq:R}
  \FOR {$\indsubpulse  = 1$ to $\numfreqs$}
    \FOR {$n = 1$ to $\numfreqs$}
        \STATE $y_{n,\indsubpulse } \gets \max_{\Bar{m} \in \{\numphases(n-1)+1, ..., \numphases n\}} x_{\Bar{m},\indsubpulse }$
        \STATE $\Bar{m}_{n,\indsubpulse } \gets \argmax_{\Bar{m} \in \{\numphases(n-1)+1, ..., \numphases n\}} x_{\Bar{m},\indsubpulse }$
    \ENDFOR
  \ENDFOR
  \STATE $\boldsymbol{Y} \gets (y_{n,\indsubpulse }) \in \mathbb{R}^{\numfreqs\times \numfreqs}$
  \STATE Apply the Hungarian algorithm to matrix $\boldsymbol{Y}$ following the steps in \cite[algorithm~1]{Senanayake22TWC} to get $\numfreqs$ elements $\hat{r}_\indsubpulse$ and their row indices $\hat{n}_\indsubpulse$, where $\indsubpulse  \in \{1,...,\numfreqs\}$
  \FOR {$\indsubpulse  = 1$ to $\numfreqs$}
    \STATE Detection of the frequency of the $\indsubpulse $-th subpulse $\hat{f}_{\indsubpulse -1}^{i} \gets f_{\hat{n}_\indsubpulse -1}$
    \STATE Detection of the phase of the $\indsubpulse $-th subpulse $\hat{\theta}_{\indsubpulse -1}^{i} \gets \frac{2\pi\lfloor(\Bar{m}_{\hat{n}_\indsubpulse ,\indsubpulse } - 1) / \numphases \rfloor}{\numphases}$
  \ENDFOR
  
 \end{algorithmic} 
 \end{algorithm} 

\subsection{Numerical Examples} \label{sec:num_comms}
\begin{figure}[t]
    \centerline{\includegraphics[width=10cm,height=7.8cm]{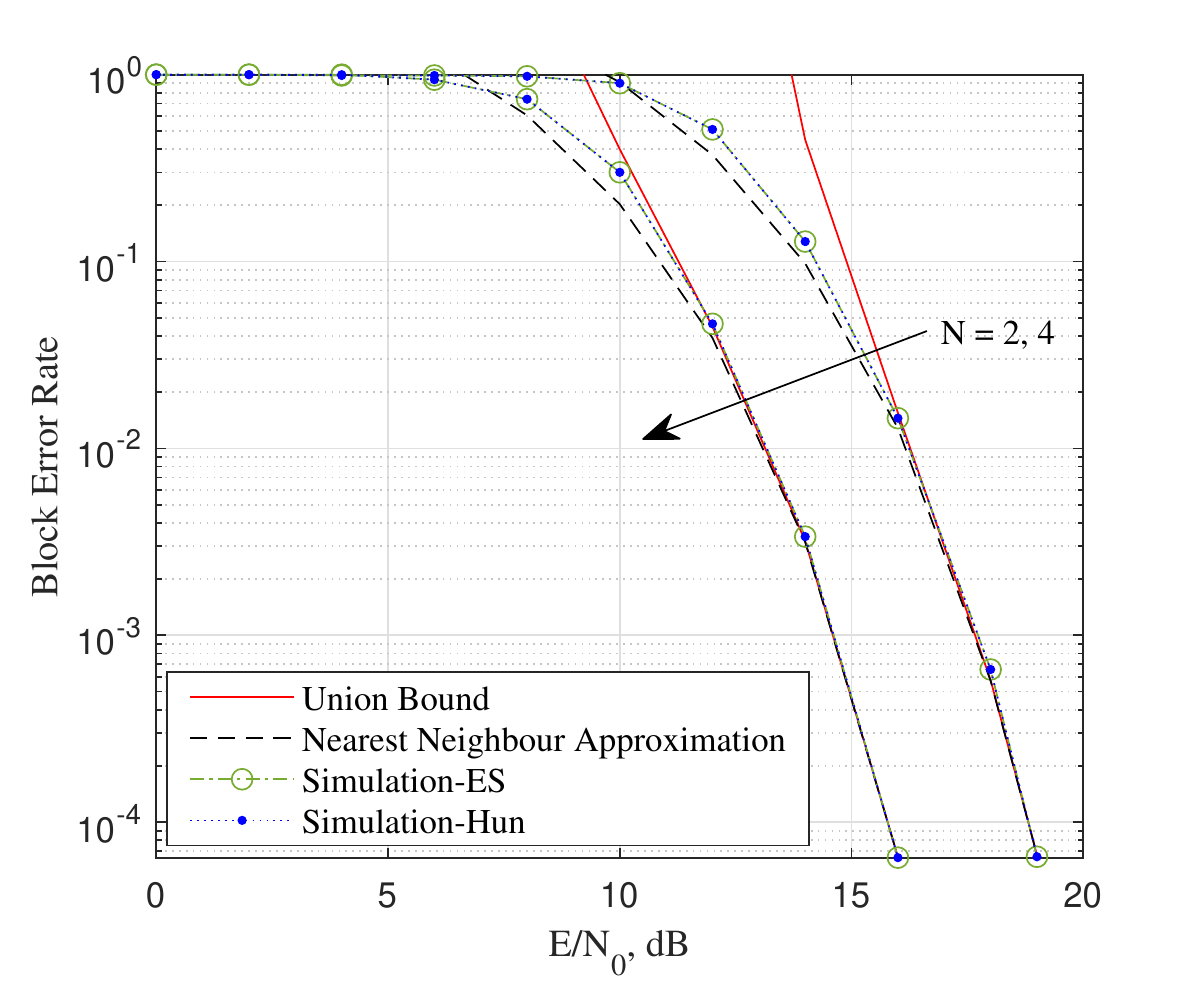}}
    \caption{The block error rate versus average received SNR for $\numfreqs = 8$, $\numphases =4$ and $N = 2,4$. The results are for the proposed baseband signalling model in an AWGN channel.
    }\label{fig:AWGN}
\end{figure}

\begin{figure}[t]
    \centerline{\includegraphics[width=10cm,height=7.8cm]{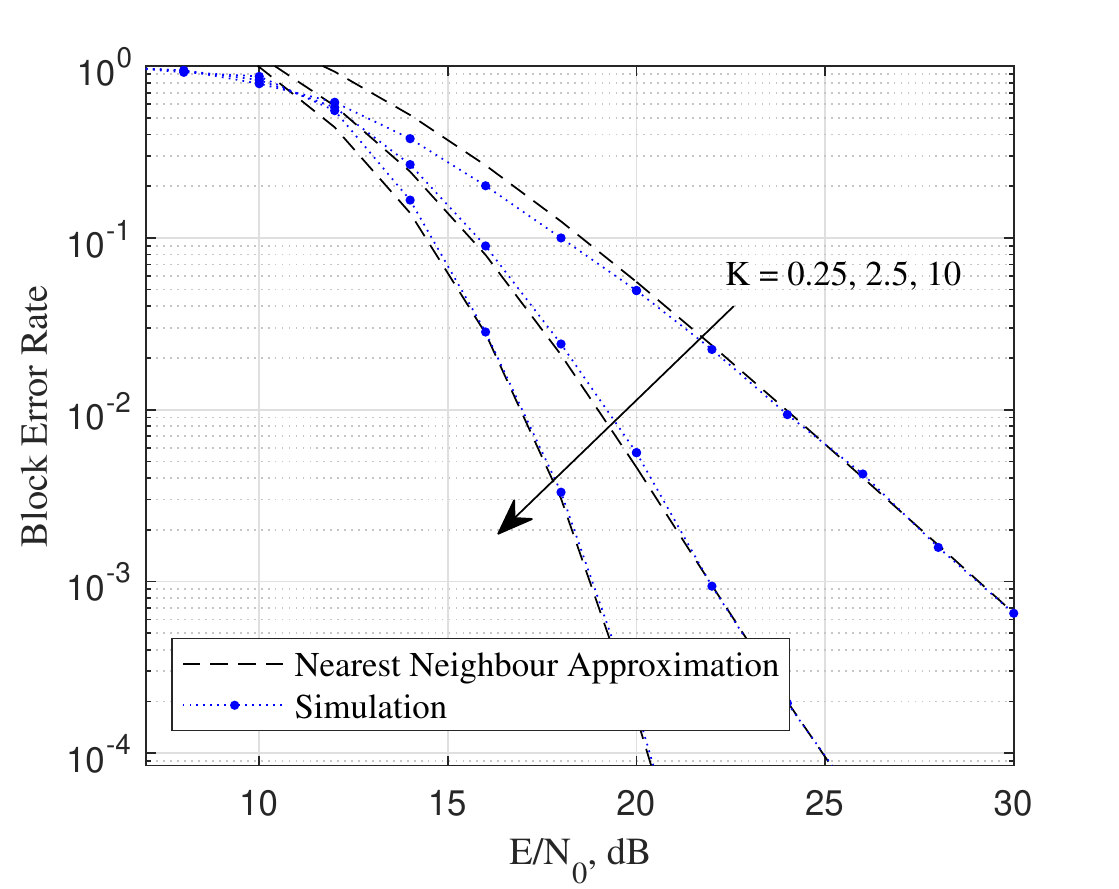}}
    \caption{The block error rate versus average received SNR for $\numfreqs = 8$, $\numphases =4$ and $N = 2$. The results are for the proposed baseband signalling model in correlated Rician fading channels with $K = 0.25, 2.5, 10$ and $\rho = 0.5$.
    }\label{fig:Rician}
\end{figure}

\begin{figure}[t]
    \centerline{\includegraphics[width=10cm,height=7.8cm]{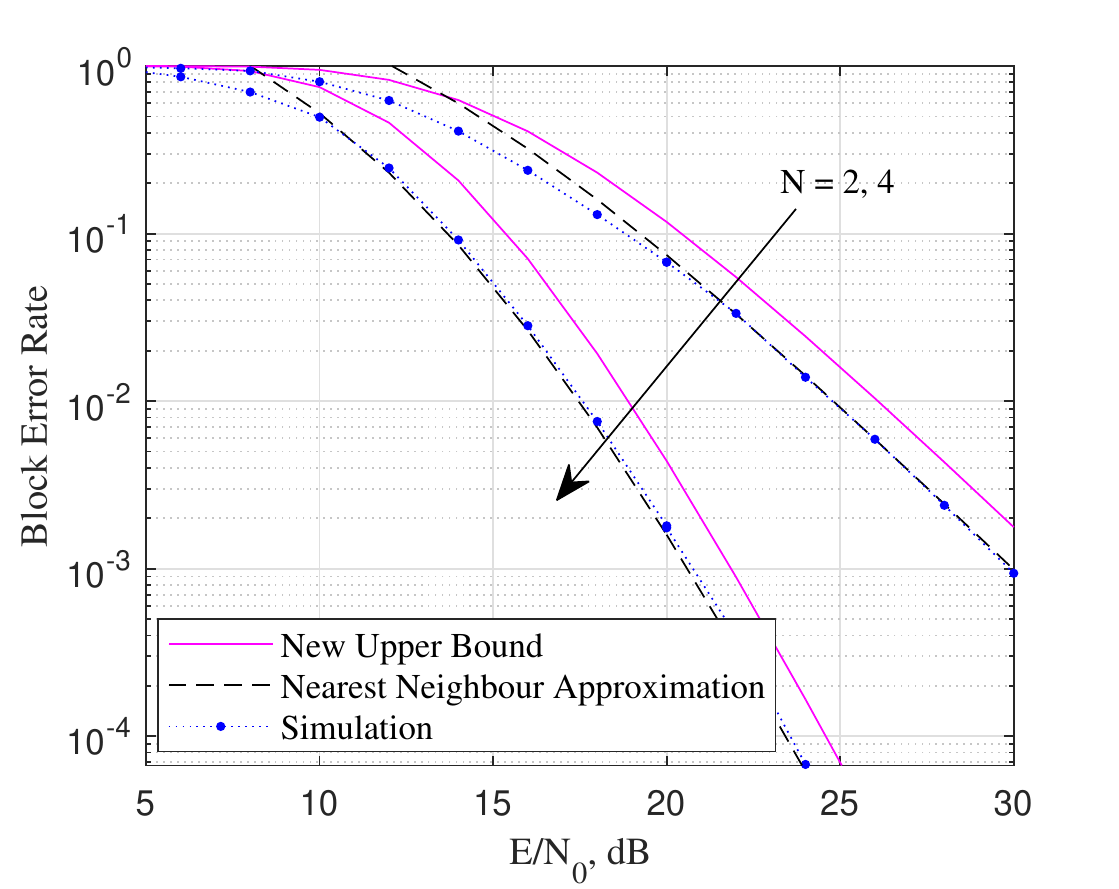}}
    \caption{The block error rate versus average received SNR for $\numfreqs = 8$, $\numphases =4$ and $N = 2, 4$. The results are for the proposed baseband signalling model in correlated Rayleigh fading channels with $\rho = 0.5$.
    }\label{fig:Rayleigh}
\end{figure}

In this section we provide numerical examples to support the theoretical analysis of the communication performance in Section \ref{sec:pe}. Fig. \ref{fig:AWGN} to \ref{fig:Rayleigh} illustrate the error probability of the waveform over different communication channel models. We mainly focus on the performance of the bounds and approximations derived in Section \ref{sec:pe}. 
   
Note that the analysis for correlated fading channels in Section \ref{sec:pe_fading} holds for all $\boldsymbol{R}_u$ but in this section we use the simple exponential correlation model. Hence, given a correlation coefficient $\rho \in [0,1]$, the $(i,j)$-th entry of $\boldsymbol{R}_u$ can be written as
\begin{equation}
\boldsymbol{R}_u(i,j) = \rho^{|i-j|}.
\end{equation}

Fig. \ref{fig:AWGN} shows the block error rate versus the average received SNR in an AWGN channel with different number of receiver antennas, $N = 2, 4$. The parameters of the waveform are kept to be $\numfreqs = 8$ and $\numphases = 4$. 
As expected, the block error rate decreases with an increase in $N$ due to the receive diversity gain. The union bounds generated using \eqref{eq:ub_awgn} always stay above the corresponding simulation results and approach the simulation curves in the high SNR region. Compared to the union bounds, the nearest neighbour approximations generated using \eqref{eq:nn_awgn} provide better approximations to the simulation results, especially in the low SNR region. 

In order to verify the optimality of the efficient receiver, we plot the error probability performance of the Hungarian algorithm based receiver and the exhaustive search based receiver, denoted as "Simulation-Hun" and "Simulation-ES", respectively. Given the same received signal, the outputs of the two receivers are exactly the same, which results in the same error performance. For the convenience of simulation, we only provide error probability results from the Hungarian algorithm based receiver in the following figures.

Fig. \ref{fig:Rician} plots the block error rate versus the average received SNR in correlated Rician fading channels with different channel factors, $K = 0.25, 2.5, 10$. We set the waveform parameters, the number of receiving antennas and the correlation coefficient to be $\numfreqs = 8$, $\numphases = 4$, $N = 2$ and $\rho = 0.5$, respectively. As has been discussed, in the fading channel the union bounds are very loose. Thus, we only show the nearest neighbour approximations using \eqref{eq:NN} with the PEP and $\alpha_{1k}^{NN}$ provided in \eqref{eq:Ric_pep} and \eqref{eq:alphaNN}, respectively. The block error rate decreases with $K$, since the LOS component is increased. The nearest neighbour approximations accurately approximate the corresponding simulation results in the high SNR region. However, the approximations become looser for smaller $K$ in the low SNR region.

Fig. \ref{fig:Rayleigh} plots the block error rate versus the average received SNR in a correlated Rayleigh fading channel with different numbers of receiver antennas, $N = 2, 4$. We set the waveform parameters and the correlation coefficient to be $\numfreqs = 8$, $\numphases = 4$ and $\rho = 0.5$, respectively. The new upper bounds in \eqref{eq:Ray_NB} are provided instead of the union bounds. The nearest neighbour approximations are generated using \eqref{eq:NN} with PEP and $\alpha_{1k}^{NN}$ provided in \eqref{eq:Ray_pep} and \eqref{eq:alphaNN}, respectively. Similar to what we observe in Fig. \ref{fig:AWGN}, the block error rate decreases with an increase in $N$. Also, the nearest neighbour approach approximates the simulation results more accurately with increasing $N$, especially in the low SNR region. The new upper bound is a tighter and more accurate performance bound on the block error rate compared to the union bound, which is not plotted in Fig. \ref{fig:Rayleigh} since the value is always larger than 1. Therefore, for the proposed signalling scheme, the new upper bound does outperform the union bound over Rayleigh fading channels.

\section{Conclusion and Future Extensions} \label{sec:conclusion}
In this paper, we consider the integration of radar sensing and communications and propose a linear stepped frequency waveform and a frequency permutation based waveform, both with phase modulation. Compared to modulating data using phase only, the randomisation in frequency tones in the second signalling scheme allows more data to be transmitted.

Focusing on the fundamental theoretical aspects, we analyse the effect of phase and frequency modulation on the radar sensing functionality and establish an important and fundamental theoretical result which shows that the phase modulation has negligible effect on the radar local accuracy. More specifically, we derive the AFs of both waveforms to analyse the impact of phase change on local accuracy and side-lobe levels. To provide quantitative measurements of the impact of phase modulation on the local accuracy, we also derive approximations to the CRLBs on delay and Doppler estimation errors based on the FIMs. We conclude that the frequency permutations result in reasonably low AF side-lobe level in average and the phase modulation has little impact on the local estimation accuracy. From the perspective of communications, we focus on analysing the block error probability of the phase modulated frequency permutation based waveform. We consider the optimal ML detection and derive the union bound and the nearest neighbour approximation on the block error rate over AWGN channels and correlated Rician fading channels. We also propose a new tighter upper bound on the block error probability, for which a closed-form expression is derived under the case of Rayleigh fading channels. 

Possible future extensions of this work includes the quantitative analysis of the effect of phase and frequency modulation on the side-lobe structure of the AF. From the communications perspective, it is desirable to extend the analysis to incorporate coding based on which a subset of possible waveforms can be selected to improve the error probability performance.



%

\appendices
\section{Derivation of the Pairwise Error Probability $P_{1k}$ over Rician fading channels} \label{sec:ricianpep}
Using \eqref{eq:rician_channel_vector} and the unitary property of $\boldsymbol{V}$, we write $\boldsymbol{h}^H\boldsymbol{h}$ as
\begin{equation}
\begin{aligned}
\boldsymbol{h}^H\boldsymbol{h} &= \left( \sqrt{\frac{K}{K+1}}\Delta^H + \sqrt{\frac{1}{K+1}} \boldsymbol{u}^H (\boldsymbol{R}_u^{1/2})^H\right)\left( \sqrt{\frac{K}{K+1}}\Delta + \sqrt{\frac{1}{K+1}} \boldsymbol{R}_u^{1/2} \boldsymbol{u} \right)\\
&= \left( \sqrt{\frac{K}{K+1}}\Delta^H \boldsymbol{V} + \sqrt{\frac{1}{K+1}} \boldsymbol{u}^H  (\boldsymbol{R}_u^{1/2})^H \boldsymbol{V}\right)\left( \sqrt{\frac{K}{K+1}}\boldsymbol{V}^H\Delta + \sqrt{\frac{1}{K+1}}\boldsymbol{V}^H \boldsymbol{R}_u^{1/2} \boldsymbol{u} \right). \label{eq:Ric_hHh1}
\end{aligned}
\end{equation}
Using \eqref{eq:eig_decomp_Ru}, we reexpress $\sqrt{\frac{1}{K+1}}\boldsymbol{V}^H \boldsymbol{R}_u^{1/2} \boldsymbol{u}$ as
\begin{equation}
 \sqrt{\frac{1}{K+1}}\boldsymbol{V}^H \boldsymbol{R}_u^{1/2} \boldsymbol{u} = \boldsymbol{D} \Tilde{\boldsymbol{u}},
\end{equation}
where $\boldsymbol{D} = \sqrt{\frac{1}{K+1}} \boldsymbol{\Omega}^{1/2}$ is diagonal and $\Tilde{\boldsymbol{u}} = \boldsymbol{V}^H \boldsymbol{u} \sim \mathcal{CN}(0,\boldsymbol{I}_N)$. Writing $\boldsymbol{D} = \text{diag}(d_1,...,d_N)$ where $d_j = \sqrt{\frac{\lambda_j}{K+1}}$, $j \in \{1,...,N\}$, we express \eqref{eq:Ric_hHh1} as
\begin{equation}
\begin{aligned}
\boldsymbol{h}^H\boldsymbol{h} &= \left( \sqrt{\frac{K}{K+1}}\Delta^H \boldsymbol{V} + \Tilde{\boldsymbol{u}}^H\boldsymbol{D}^H\right)\left( \sqrt{\frac{K}{K+1}}\boldsymbol{V}^H\Delta + \boldsymbol{D} \Tilde{\boldsymbol{u}}\right)\\
& = \sum_{j = 1}^N \left| \sqrt{\frac{K}{K+1}} (\boldsymbol{V}^H\Delta)_j + d_j \Tilde{\boldsymbol{u}}_j \right|^2\\
& = \sum_{j = 1}^N \left(d_j \Re\left\{\Tilde{\boldsymbol{u}}_j\right\} + \sqrt{\frac{K}{K+1}}\Re\left\{(\boldsymbol{V}^H\Delta)_j\right\} \right)^2 + \left(d_j \Im\left\{\Tilde{\boldsymbol{u}}_j\right\} + \sqrt{\frac{K}{K+1}}\Im\left\{(\boldsymbol{V}^H\Delta)_j\right\} \right)^2, \label{eq:Ric_hHh2}
\end{aligned}
\end{equation}
where $\Tilde{\boldsymbol{u}}_j$ is the $j$-th element of $\Tilde{\boldsymbol{u}}$, $(\boldsymbol{V}^H\Delta)_j$ is the $j$-th element of $\boldsymbol{V}^H\Delta$ and $\Im\{\cdot\}$ denotes the imaginary part of the argument. We can further rewrite \eqref{eq:Ric_hHh2} as
\begin{equation}
\boldsymbol{h}^H\boldsymbol{h} = \sum_{j=1}^{2N} \nu_j(W_j-w_j)^2, \label{eq:Ric_hHh}
\end{equation}
where $W_j \sim \mathcal{N}(0,1)$, 
\begin{equation}
\begin{aligned}
\nu_j = \left\{\begin{aligned}
&\frac{\lambda_j}{2(K+1)}, & 1 \leq j \leq N\\
&\frac{\lambda_{j-N}}{2(K+1)}, & N+1 \leq j \leq 2N,
\end{aligned}
\right.  
\end{aligned} \label{eq:nuj}
\end{equation}
and
\begin{equation}
\begin{aligned}
w_j = \left\{\begin{aligned}
&-\sqrt{\frac{2K}{\lambda_j}}\Re\left\{(\boldsymbol{V}^H\Delta)_j\right\}, & 1 \leq j \leq N\\
&-\sqrt{\frac{2K}{\lambda_{j-N}}}\Im\left\{(\boldsymbol{V}^H\Delta)_{j-N}\right\}, & N+1 \leq j \leq 2N.
\end{aligned}
\right.  
\end{aligned} \label{eq:wj}
\end{equation}
With the expression of $\boldsymbol{h}^H\boldsymbol{h}$ in \eqref{eq:Ric_hHh}, we can now derive $P_{1k}$ over the correlated Rician fading channel. Using \eqref{eq:pik} and conditioning on the distribution of $\boldsymbol{h}^H\boldsymbol{h}$, we express $P_{1k}$ as
\begin{equation}
\begin{aligned}
P_{1k} &= \mathbf{E}_{\boldsymbol{h}^H\boldsymbol{h}}\left[ \Pr\left[ Z > \frac{\sqrt{\boldsymbol{h}^H\boldsymbol{h}}}{\alpha_{1k}} \mid \boldsymbol{h}^H\boldsymbol{h}\right] \right]\\
&= \mathbf{E}_{\boldsymbol{h}^H\boldsymbol{h}}\left[Q\left( \frac{\sqrt{-\boldsymbol{h}^H\boldsymbol{h}}}{\alpha_{1k}} \right)\right], \label{eq:Ric_p1k1}
\end{aligned}
\end{equation}
where $\mathbf{E}_{X}[\cdot]$ is the expectation of the argument with respect to the distribution of $X$. Using Craig's formula to reexpress the Gaussian Q-function, we write \eqref{eq:Ric_p1k1} as
\begin{equation}
P_{1k} = \frac{1}{\pi} \int_0^{\pi/2} \mathbf{E}_{\boldsymbol{h}^H\boldsymbol{h}}\left[ \exp \left(\frac{\boldsymbol{h}^H\boldsymbol{h}}{2 \alpha_{1k}^2 \sin^2 \theta} \right) \right] d\theta.
\end{equation}
Substituting $\boldsymbol{h}^H\boldsymbol{h}$ with \eqref{eq:Ric_hHh} we obtain
\begin{equation}
P_{1k} = \frac{1}{\pi} \int_0^{\pi/2}  \prod_{j=1}^{2N} \mathbf{E}_{W_j} \left[\exp \left( -\frac{\nu_j (W_j-w_j)^2}{2\alpha_{1k}^2\sin^2\theta}  \right)\right] d\theta. \label{eq:Ric_p1k2}
\end{equation}
Since $W_j \sim \mathcal{N}(0,1)$, we rewrite the expectation in \eqref{eq:Ric_p1k2} as
\begin{equation}
\begin{aligned}
\mathbf{E}_{W_j} \left[\exp \left( -\frac{\nu_j(W_j-w_j)^2}{2\alpha_{1k}^2\sin^2\theta}  \right)\right] &= \int_{-\infty}^{\infty} \frac{1}{\sqrt{2\pi}} \exp \left( -\frac{\nu_j(W_j-w_j)^2}{2\alpha_{1k}^2\sin^2\theta}  - \frac{W_j^2}{2}\right) d W_j\\
&= \sqrt{\frac{\alpha_{1k}^2\sin^2\theta}{\nu_j + \alpha_{1k}^2\sin^2\theta}}\exp \left( -\frac{\nu_j w_j^2}{2(\nu_j+\alpha_{1k}^2\sin^2\theta)}\right). \label{eq:Ric_E}
\end{aligned}
\end{equation}
Substituting \eqref{eq:Ric_E}, \eqref{eq:nuj} and \eqref{eq:wj} into \eqref{eq:Ric_p1k2} and using some straight forward mathematical manipulations, we can obtain the expression of $P_{1k}$ in \eqref{eq:Ric_pep}.

\section{Derivation of the Conditional Pairwise Error Probability $\Tilde{P}_{1k}$} \label{sec:conditioned_pik}
Using the results in \eqref{eq:pik} and \eqref{eq:conditioned_pik}, the PEP conditioned on $\boldsymbol{h}^H\boldsymbol{h} \geq \gamma$ can be expressed as
\begin{equation}
\Tilde{P}_{1k} = \Pr[\sqrt{\boldsymbol{h}^H\boldsymbol{h}}  < \alpha_{1k}Z \mid \boldsymbol{h}^H\boldsymbol{h} \geq \gamma], \label{eq:con_P1k1}
\end{equation}
where $Z \sim \mathcal{N}(0,1)$. By conditioning on $\boldsymbol{h}^H\boldsymbol{h}$, \eqref{eq:con_P1k1} can be written as
\begin{equation}
\begin{aligned}
\Tilde{P}_{1k} &= \int_{\gamma}^{\infty} Q\left( \frac{\sqrt{x}}{\alpha_{1k}} \right) f_{\boldsymbol{h}^H\boldsymbol{h} \mid \boldsymbol{h}^H\boldsymbol{h} \geq \gamma}(x) dx\\
&= \int_{\gamma}^{\infty} Q\left( \frac{\sqrt{x}}{\alpha_{1k}} \right) \frac{f_{\boldsymbol{h}^H\boldsymbol{h}}(x)}{\Pr[\boldsymbol{h}^H\boldsymbol{h} \geq \gamma]} dx, \label{eq:con_P1k2}
\end{aligned}
\end{equation}
where $f_{\boldsymbol{h}^H\boldsymbol{h}}(x)$ is the probability density function of $\boldsymbol{h}^H\boldsymbol{h}$. By applying Craig's formula to \eqref{eq:con_P1k2}, we obtain the expression in \eqref{eq:conditioned_pik}.




\ifCLASSOPTIONcaptionsoff
  \newpage
\fi



%



\bibliographystyle{IEEEtran}

%








\end{document}